\documentclass[twocolumn,english,groupedaddress, superscriptaddress,aps,pre,10pt,floatfix]{revtex4-1}
\usepackage[T1]{fontenc}
\usepackage[utf8]{inputenc}
\usepackage{amsmath,amsthm,mathtools,amssymb}

\usepackage{appendix}
\usepackage{listings}

\usepackage{units}

\usepackage{url}
\PassOptionsToPackage{hyphens}{url}
\usepackage{hyperref}

\usepackage[usenames,dvipsnames,svgnames,table]{xcolor}
\definecolor{blue}{rgb}{0,0,1}

\definecolor{dgreen}{rgb}{0.0,0.5,0.0}

\definecolor{dred}{rgb}{0.5,0,0}

\definecolor{dyellow}{rgb}{0.75,0.75,0}

\definecolor{lightBlue}{rgb}{0.,0.5,0.5}

\usepackage[colorinlistoftodos]{todonotes}

\renewcommand{\d}{\mathrm{d}}

\begin{document}

%\title{Stochastic model that generates synthetic power-grid frequency data}
%\title{Modelling the dynamics of power-grid frequency fluctuations from data}
%\title{Data-driven synthetic of the power-grid frequency model}
%\title{Data-driven model of the mains frequency}
\title{Data-driven model of the power-grid frequency dynamics}

\author{Leonardo Rydin Gorj\~ao}
\email{l.rydin.gorjao@fz-juelich.de}
\affiliation{Forschungszentrum J\"ulich, Institute for Energy and Climate Research - Systems Analysis and Technology Evaluation (IEK-STE), 52428 J\"ulich, Germany}
\affiliation{Institute for Theoretical Physics, University of Cologne, 50937 K\"oln, Germany}

\author{Mehrnaz Anvari}
\affiliation{Max--Planck Institute for the Physics of Complex Systems (MPIPKS), 01187 Dresden, Germany}

\author{Holger Kantz}
\affiliation{Max--Planck Institute for the Physics of Complex Systems (MPIPKS), 01187 Dresden, Germany}

\author{Christian Beck}
\affiliation{School of Mathematical Sciences, Queen Mary University of London, London E1 4NS, United Kingdom}

\author{Dirk Witthaut}
%\email{d.witthaut@fz-juelich.de}
\affiliation{Forschungszentrum J\"ulich, Institute for Energy and Climate Research - Systems Analysis and Technology Evaluation (IEK-STE), 52428 J\"ulich, Germany}
\affiliation{Institute for Theoretical Physics, University of Cologne, 50937 K\"oln, Germany}

\author{Marc Timme}
\thanks{Contributed equally}
%\email{marc.timme@tu-dresden.de}
\affiliation{Chair for Network Dynamics, Center for Advancing Electronics Dresden (cfaed) and Institute for Theoretical Physics, Technical University of Dresden, 01062 Dresden, Germany}
\affiliation{Network Dynamics, Max Planck Institute for Dynamics and Self-Organization (MPIDS), 37077 G\"ottingen, Germany}

\author{Benjamin Sch\"afer}
%\email{b.schaefer@qmul.ac.uk}
\thanks{Contributed equally}
\affiliation{School of Mathematical Sciences, Queen Mary University of London, London E1 4NS, United Kingdom}
\affiliation{Chair for Network Dynamics, Center for Advancing Electronics Dresden
(cfaed) and Institute for Theoretical Physics, Technical University
of Dresden, 01062 Dresden, Germany}
\affiliation{Network Dynamics, Max Planck Institute for Dynamics and Self-Organization (MPIDS), 37077 G\"ottingen, Germany}

%formalities: http://ieeeaccess.ieee.org/frequently-asked-questions/
%Typical articles had length~ 6-10 pages

\begin{abstract}
The energy system is rapidly changing to accommodate the increasing number of renewable generators and the general transition towards a more sustainable future.
%Simultaneously, new business models and market designs are proposed to stabilise the power grid and its frequency.
Simultaneously, business models and market designs evolve, affecting power-grid operation and power-grid frequency.
%Simultaneously, new  business models and regulatory frameworks are proposed, which affect the power grid operation and its frequency.
Problems raised by this ongoing transition are increasingly addressed by transdisciplinary research approaches, ranging from purely mathematical modelling to applied case studies.
These approaches require a stochastic description of consumer behaviour, fluctuations by renewables, market rules, and how they influence the stability of the power-grid frequency.
Here, we introduce an easy-to-use, data-driven, stochastic model for the power-grid frequency and demonstrate how it reproduces key characteristics of the observed statistics of the Continental European and British power grids.
We offer executable code and guidelines on how to use the model on any power grid for various mathematical or engineering applications.
%\sout{A more thorough examination of the power grid statistical "nature"(?) demands/request a deepened look into the stochatic process governing power-grid frequency.}
%\sout{Transdisciplinary interest in power grid research is growing, bringing solutions to \sout{surmount the} the problems entailed in the increasing number of renewable generators into the energy system. }
%\sout{The increasing renewable generation brings forth new problems to the stability of power grids. To overcome the ...}
\end{abstract}

\maketitle

% \section*{ToDo:}
% \begin{itemize}
% 	\item @all: [Politics] Decide where and how to publish: PSCC conference or IEEE access. \textbf{Possible compromise:
%     Submit full paper early September to IEEE, if editors reject, submit to conference\\
%     + Ask PSCC conference whether non-Elsevier journal is an option instead of EPSR special issue}
%     \item Schedule: PSCC deadline: 01.10.2019
% 	\item @Submitting Author: Update cross citations PRE and IEEE after arxiv submission, preliminary titles and author lists are used in Anvari2019 and Gorjao2019; check for too much redundancy
% 	\item @Submitting Author: Move "Supplemental Material" section with code into new "Supplemental Material" file before submission.
% \end{itemize}
\section{Introduction}

%Introduce the research field, commenting on the  challenges of new market designs, business and microgrid models
The energy system is currently undergoing a rapid transition towards a more sustainable future.
Greenhouse gas emissions are reduced by implementing distributed renewable-energy sources at ever growing rates in the world \cite{sawin2018renewables}.
Simultaneously, new policies, technologies, and market structures are being implemented in various regions in the energy systems \cite{rodriguez2014business}.
These new market structures are not necessarily benefiting the stability of the power grid: A control power shortage in the German grid in June 2019 was potentially caused by unknown traders exploiting the energy market structure \cite{Spiegel2019}. 

%quickly developing field of energy research
The field of energy research itself is quickly developing and attracting researchers from various disciplines working towards new control systems, new market models, and new technologies every year \cite{panwar2011role,tuballa2016review}.
Regardless of the specific aspect of the energy system, one element remains unchanged: The electrical power system and the stability of its frequency are  critical for a stable operation of our society \cite{Obama2013}.
    
%Explain how and why the mains frequency is an important observable for the state of the power grid
The power-grid (mains) frequency dynamics mirrors the balance of supply and demand of the power grid: An excess of generation leads to an increased frequency and a shortage of generation leads to a reduced frequency value.
The power grid is stabilised by controlling the frequency and maintaining it at a nominal frequency \cite{Kundur1994}.  
But the task of maintaining a set frequency across an entire power-grid system is not a simple one: systems vary in size and structure, energy sources are possibly volatile in their output, as for example are wind or photo-voltaic generators \cite{Anvari2016,wolff2019}, and the dispatch of electrical energy and market activity have an impact on the overall dynamics.

%Motivate necessity of frequency model
Understanding the intricacies of the frequency dynamics becomes of great importance, both to control the current power grid \cite{Kundur1994, Rohden2012} but also for implementing real-time pricing schemes \cite{Walter2014,Schaefer2015} or smart grids in the future \cite{Fang2012}. 
Solid estimates of fluctuations are essential for example when dimensioning back-up or control options, such as determining the capacity of batteries or other energy storage to balance periods with highly fluctuating demand or times without renewable generation \cite{weber2018wind}.
Similarly, when establishing new power grid types, such as smart grids with potentially novel electricity market structure, the market design should ideally support the stability of the grid. 
 
%Review previous stochastic and deterministic modelling of  power grids
While both the power-grid frequency dynamics and the stochastic nature of the power-grid frequency have been intensely studied, we require a better understanding of the interaction of frequency dynamics with both stochastic fluctuations and market behaviour.
The dynamics of the power-grid variables, including frequency, voltage, reactive power, etc., may be modelled with arbitrary complexity based on various models \cite{Kundur1994,Filatrella2008,Rohden2012,Nishikawa2015,schmietendorf2017,bottcher2019}.
Simultaneously, stochastic modelling of fluctuations within the power grid \cite{schmietendorf2017,Haehne2018} still often uses Gaussian noise models \cite{Zhang2010,Fang2012,Schaefer2017}, while non-Gaussian statistics \cite{Anvari2016,Schaefer2017a} as well as deterministic events caused by trading \cite{Weisbach2009} are rarely included. 

%Previous models of power-grid frequency forecasts and predictions
Existing literature often focuses on inverter control \cite{scoltock2015model} or the power interface between grid layers \cite{dong2012frequency}.
Even forecasting is mostly done for electricity consumption \cite{tso2007predicting} or for renewable generation, such as solar generators \cite{sharma2011predicting}.
In contrast, models that predict or even give stochastic characteristics of the power-grid frequency are very rare \cite{tchuisseu2017}.

%Overview and structure of the article
Here, we propose an accessible and easy-to-use stochastic model that seeks to describe the dynamics of the power-grid frequency in a reduced framework combining stochastic and deterministic factors acting on the power-grid frequency.
We focus on the intermediate time scale of several seconds to few hours, leaving very short or very long time scale for future work.
Simultaneously, our modelling approach balances the benefits of realistic case studies, generally applicable and abstract stochastic models as well as application-oriented data-driven approaches. 

We first review the factors influencing the power-grid frequency dynamics, based on frequency recordings from European grids.
Next, we introduce a general stochastic model and discuss three particular cases of how the model may be implemented. 
For each case we estimate the system parameters, such as control strength and noise amplitude using stochastic theory and data-driven approaches.
%These model cases range from a very basic and mathematically easily tractable case to a much more applied and real-world-focused case. 
We compare the frequency statistics of the models with real-world measurements to showcase how they reproduce characteristic features.
Overall, our modelling approach is very flexible and easily applicable to many different power grids and could be used for planning purposes, e.g. when setting security operational limits or designing markets. We provide executable code for the model in the supplementary material.

\section{Factors impacting the power-grid frequency}
%Introduction on what influences the grid
To construct a model describing the intermediate time scale dynamics and characteristics of the power-grid frequency, we must first recall the nature and the intricate details of the power-grid frequency dynamics, both deterministic and stochastic, as we observe them in frequency trajectories \cite{Transnet}, see Fig.~\ref{fig:Trajectory}.

The power-grid frequency is not following a simple Gaussian process but displays heavy tails and regular correlation peaks, see Fig.~\ref{fig:Histogram_Autocorrelation_Data} and \cite{Schaefer2017a,Schafer2018b,Anvari2019} for more detailed analysis. To get a better understanding of the different factors impacting the grid frequency, we give an overview of these: First, we review the innate and humanly devised control systems, continue with the market and power dispatch design and close the section with a stochastic description of the noise acting on the power grid.

%different control methods and time scales
\subsection{The fundamental control schemes}
The power supply of the grid is designed so that the frequency of the alternating current is kept steadily at a fixed nominal value, i.e., $50$~Hz in Europe and many parts of the world, or $60$~Hz in the Americas, Southern Japan and some other regions.
All power plants in a given synchronous region, such as the Continental European grid or the Eastern Interconnection of North America, are designed to operate at this reference frequency.
The electrical frequency of e.g. $50$~Hz corresponds to large mechanical generators rotating in synchrony at this frequency (or integer multiples of it) across the entire region.
How is this frequency kept fixed when facing fluctuations or larger disturbances?

%generator disconnects: inertial, primary and secondary controls activate
Suppose a large generator disconnects from the grid while the power demand in the region stays constant.
The missing energy cannot be drawn from the grid itself, as it cannot store any energy directly \cite{Machowski2011}.
Instead, power is first provided by inertial energy until  primary, secondary, and potentially tertiary control set in to ensure the provision of the missing power \cite{Machowski2011}.
%energy is first drawn from the inertia of the rotating generators, and next, primary, secondary, and potentially tertiary control set in to ensure the provision of the missing power \cite{Machowski2011}.
%
In the first moments after the disturbance, the missing power is drawn from the kinetic energy of the large rotating machines.
Their kinetic energy is converted into electrical energy and the generators are slowed down, thereby reducing the overall frequency in the grid.
This \emph{inertial response} ensures the system does not drift off from its designed nominal frequency too rapidly and smoothens any disturbances.
Nevertheless, the generators continue to slow down. 
Moments later, \emph{primary control} activates: Dedicated power plants, and recently also battery stacks  \cite{oudalov2007optimizing}, measure the deviation of the frequency from the reference and insert additional power into the grid proportional to the frequency deviation.
This power influx prevents a further decrease of the frequency and stabilises it at a fixed but lower frequency, which is not desired for operation, as any further problems might cause the frequency to leave the stable operational limits \cite{Machowski2011, Kundur1994}.
While the primary control compensates for the missing power, the kinetic energy of the rotors is still lower than initially and thereby the frequency is not at the reference value.
To restore the frequency back to the reference frequency an integrative control, \emph{secondary control}, is necessary.
A few minutes after the disturbance, this control fully restores the energetic state and the grid is brought back into a new stable state at its nominal frequency (i.e., $50~$Hz or $60~$Hz, depending on the grid in question).
On even longer time scales of potentially hours, \emph{tertiary control}, often operated manually, sets in \cite{Wood2012}. As this tertiary control sets in, primary and secondary control can be reduced to become available for further control actions.

Here, we focus on the effects of inertia, as well as primary (proportional) and secondary (integrative) control in our synthetic model. 
The time scales of these three controls are significantly different, and they functionally react to deviations of different variables of the system: Where primary control stabilises the grid based on the frequency deviations of the system, the secondary control balances the total power to ensure stability based on an integral of the frequency, i.e., an angle.

As a recent challenge, the replacement of conventional power generators with renewable generators reduces the overall system inertia \cite{Milano2018} and thereby makes complementary control mechanisms or virtual inertia increasingly important \cite{Beck2007}.

%Dispatch, market and trading mechanisms
\subsection{Electricity dispatch and market}

%Market design overview: Start with smoothly changing demand (due to aggregation of millions of consumers), then motivate the need for regular trading actions to match supply and demand, leading to a sawtooth function (Fig.~\ref{fig:DemandVsGeneration})
While the control schemes keep the frequency close to the reference for small and unforeseen changes of supply and demand, an electricity market has been established to coordinate longer-term power dispatches dealing with large and predictable variations.

%smooth demand changing overtime
The effective demand acting on the power grid is the aggregation of millions of consumers throughout the synchronous region. This aggregated demand is continuously changing over time since consumption during the day tends to be higher then during the night and industrial activities during the week lead to higher consumption than during the weekends \cite{Machowski2011}. 

%necessity to dispatch power via regular trading actions
The continuously changing demand has to be met with sufficient supply of electrical power in the same synchronous grid. 
Therefore, power plant operators have to adjust their generation according to the needs of the consumers.
While some power plants, such as gas turbines, can ramp their generation up or down very fast, other plants, such as coal or nuclear power plants, require more time and therefore prefer to commit generation for longer time periods \cite{Kundur2004, Wood2012}. Demand response schemes, where consumers shift their demand to periods of higher generation, bring additional flexibility to the grid \cite{Palensky2011}.

%Explain market structure to determine who is supplying the power
To reach an economic optimum on who is supplying and when, power-plant operators bid on spot markets to offer power generation \cite{Wood2012}.
This includes a \emph{day-ahead} market to fulfil the expected power demand, and an \emph{intra-day} market acting on time scales of few hours to several minutes, to balance short-term mismatches, amongst other \cite{Kovacevic2013}.
This bidding on the market takes place in discrete time-slots: Any power provided by one operator is provided for a fixed interval, e.g. one hour, half an hour, or 15 minutes, as is often the case, such as in the European Energy Exchange (EEX) \cite{NationalAcademiesofSciences2016}. 

%Lead to sawtooth function of 15 minute dispatch actions vs. smooth demand
An important consequence of the fixed intervals of generation is that it does not perfectly fit the smooth demand curve. If we approximate a monotonically increasing demand function (such as during the early morning hours) with a step function assuming the mean for a given time interval, we will initially overestimate the demand, which is still growing. After some time, supply and demand perfectly match but then the demand surpasses the supply again. This leads to the balance between supply and demand being approximately a sawtooth function, see Fig.~\ref{fig:DemandVsGeneration}.

\begin{figure}
\includegraphics[width=0.99\linewidth]{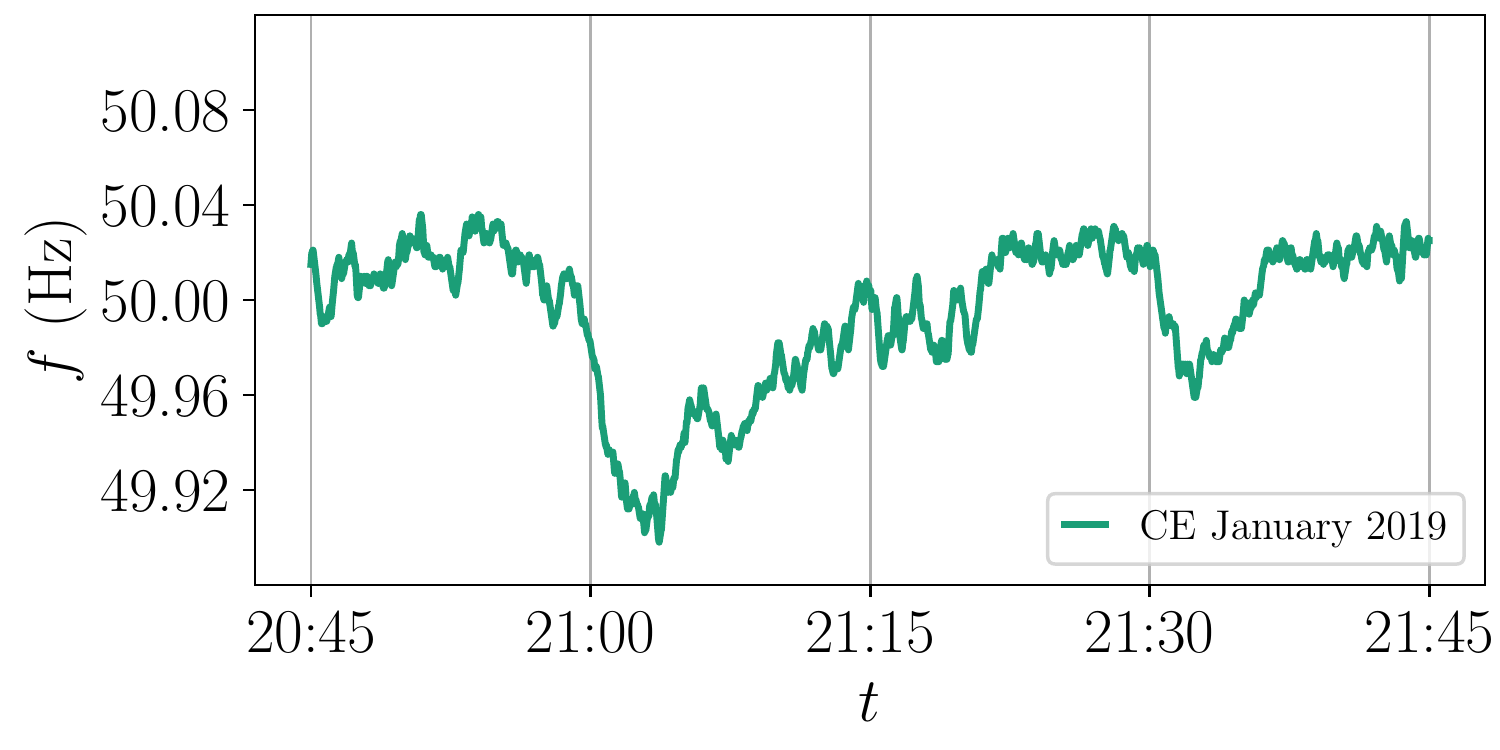}
\caption{{\bf The frequency dynamics is influenced by both stochastic and deterministic aspects.} The trajectory of the power-grid frequency is substantially influenced by stochastic effects, as seen by the erratic motion. In addition, we observe deterministic behaviour: Every $15$ minutes (vertical lines) the frequency abruptly decreases and then slowly trends upwards for the next $15$ minutes.
The plot uses the TransNetBW data \cite{Transnet} from the European Central power grid CE, from January, 10$^{\mathrm{th}}$ 2019, $20{:}45$ to $21{:}45$.}\label{fig:Trajectory}
\end{figure}

\begin{figure*}
\centering
\includegraphics[width=0.99\linewidth]{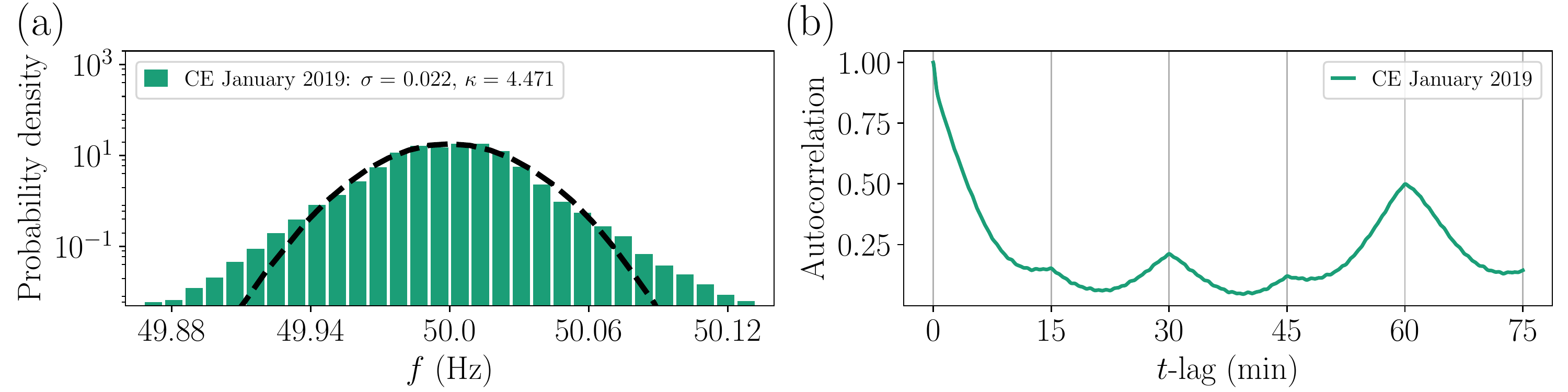}
\caption{{\bf The power-grid frequency is heavy-tailed and has regular correlation peaks.} {\textbf{(a)}} The frequency histogram displays heavy tails, which are quantified by a kurtosis $\kappa$ that is much larger than the Gaussian value of $\kappa_\text{Gaussian} = 3$. Consistently, the best-fitting Gaussian distribution (dashed line) does not capture the tails. 
{\textbf{(b)}} The autocorrelation function of the grid frequency decays exponentially within the first minutes, which is a typical behaviour for many stochastic processes \cite{Gardiner1985}. In addition, the autocorrelation peaks every 15 minutes due to trading activity.
The plots use the TransNetBW data from January 2019 \cite{Transnet}.}
\label{fig:Histogram_Autocorrelation_Data}
\end{figure*}

% observing trading effects in the data
Indeed, we also observe the consequences of the intervals when analysing the frequency trajectory \cite{Weisbach2009} or its autocorrelation in the Continental European grid.
The frequency displays regular surges and sags approximately every 15 minutes, where the supply updates to the new demand interval.
At full hours these effects are more pronounced since the total dispatch and trading volume is higher at full hours compared to other 15 minute intervals \cite{Weissbach2009}.
Not only the frequency trajectory displays these jumps and sags, see Fig.~\ref{fig:Trajectory}, but also the autocorrelation function of the power-grid frequency $c(\Delta t)$ reveals distinct peaks at $15$, $30$, $45$ and $60$ minutes, see Fig.~\ref{fig:Histogram_Autocorrelation_Data} and  \cite{Schaefer2017a,Schafer2018b}.

We will include the market influence by employing a deterministic power-mismatch model in our stochastic model. But more importantly, we can extract vital information by observing this phenomenon, as we will highlight below.

\subsection{Noise}\label{sec:Stochastic dynamics}

So far, we have introduced the two deterministic elements of our model: Control in the form of inertia, primary and secondary control, and electricity trading occurring at fixed times. We are only missing the stochastic element of the model, i.e., the \emph{noise} acting on the system.
Noise here is meant as any form of stochastic fluctuation. Its sources are plentiful, ranging from demand fluctuations \cite{gonzalez2007forecasting,Palensky2011} to intermittency in the renewable generators \cite{Milan2013, Anvari2016}, thermal fluctuations, and others, many of which are typically unknown \cite{Schaefer2017a}.
However, the precise origin of the noise is not essential for our modelling approach.
In fact, we only observe the cumulative effect of the noise in how it influences the power-grid frequency, regardless whether it originates from local disturbances or system-wide variations. Aggregating all sources of noise allows it to be handled as a stochastic process, see also \cite{Anvari2019} for more details.
%A more in-depth study of the fluctuations of the grid frequency is presented in \cite{Anvari2019}. 

%Motivate why we use Gaussian noise and justify it via central limit theorem
As a first approximation for the noise, we will assume white Gaussian noise, based on two important observations.
First,  Gaussian noise arises naturally in many settings due to the \emph{Central Limit Theorem}.
In its simplest form it states that the sum of randomly drawn numbers, in our case the aggregation of renewable, demand and any other form of fluctuation, approximates a Gaussian distribution if sufficiently many contributions are summed up \cite{Gardiner1985}. 
Second, we note that non-Gaussian frequency distributions can easily be described by super-imposed Gaussian distributions, following \emph{superstatistics} \cite{Schaefer2017a,Beck2003,Beck2007}, where parameters, such as the standard deviation change over time.
Moreover, the above mentioned trading intervals are known to contribute significantly to these tails \cite{Schafer2018b}.

%Discuss non-Gaussian noise and that they can be easily included
If so desired, employing another form of noise is left open in the model, without any fundamental change of the model itself.
There are plenty of non-Gaussian sources of noise impacting the power grid, such as jump noise from solar panels \cite{Anvari2016} or turbulence from wind turbines \cite{Milan2014, Haehne2018}.
Instead of Gaussian noise, we could include for example non-Gaussian effects via Lévy-stable distributions or $q$-Gaussian distributions \cite{Beck2005,Beck2007}.

\section{Data-driven model}
Now, we formulate a simple dynamical model for the frequency dynamics that includes all factors influencing the power-grid frequency. First, we present the model and explain how the above-mentioned factors enter the model. We then discuss special cases of how some parameters could be set as constants or as time-dependent. We close the section by proving the theory to estimate the parameters of the model.

For simplicity, we do not use the frequency $f$ as the variable, but the bulk angular velocity $\omega=2\pi \left(f-f_\text{ref}\right)$, with reference frequency $f_\text{ref}=50$ or $60$~Hz, i.e., we move into the rotating reference frame. In this frame, the dynamics of the angular velocity $\omega$ and the bulk angle $\theta$ may be modelled in an aggregated swing equation \cite{Ulbig2014} as

\begin{equation}
    \begin{aligned}\label{eq:model}
    \frac{\d \theta}{\d t}  & = \omega,\\
    M \frac{\d \omega}{\d t} & = -c_{1}\omega-c_{2}\theta+ \Delta P+\epsilon\xi.
    \end{aligned}
\end{equation}
The factor $M$ gives the inertial constant of the system and sets the time scale it reacts to changes. For simplicity, we absorb it in the remaining constants and set $M=1$ in the following, i.e., $c_1 \rightarrow c_1/M$, $c_2 \rightarrow c_2/M$,$\Delta P \rightarrow \Delta P/M$ and $\epsilon \rightarrow \epsilon$.

\begin{figure*}
\centering
\includegraphics[width=0.99\linewidth]{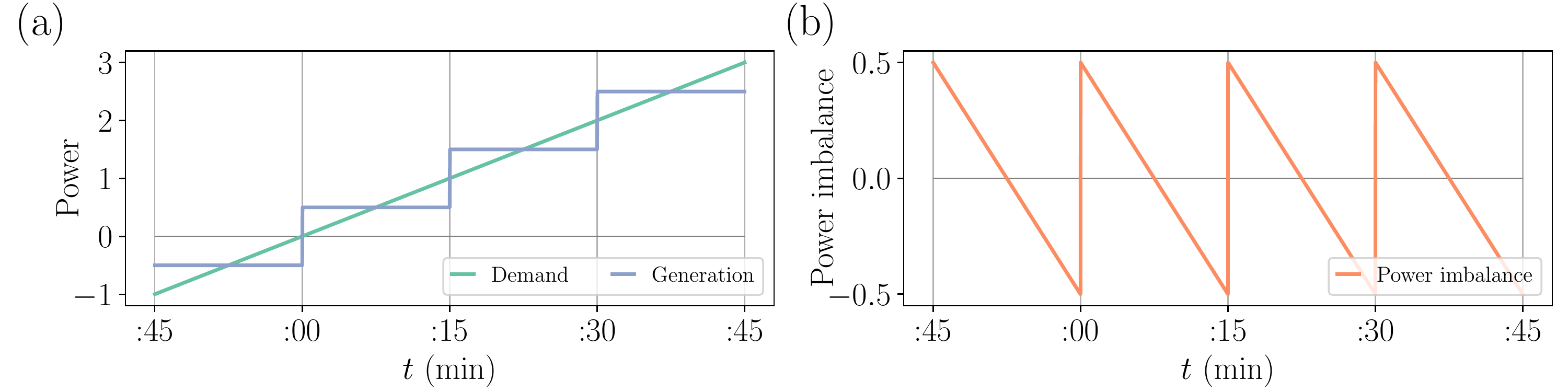}
\caption{{\bf The effective power balance approximates a sawtooth function over time.} We schematically depict the interplay between generation, demand and the resulting power imbalance:  {\textbf{(a)}} While the demand curve is approximately smooth, the scheduled generation approximates the curve using step functions. {\textbf{(b)}} The resulting power balance is approximately a sawtooth function with jumps upwards and ramps downwards if the demand rises and ramps upwards and jumps downwards if the demand decreases. Here, we display all jumps with the same height for simplicity. In our model, we use different jump heights of the Heaviside and thereby also of the sawtooth function for hourly, half- or quarter-hourly jumps.}
\label{fig:DemandVsGeneration}
\end{figure*}

The term $-c_1 \omega$ models primary control and general damping acting on the system \cite{Machowski2011, Filatrella2008}. The larger the deviation from the nominal frequency, i.e., the larger $\omega$, the larger the damping and control force.

The expression $-c_2 \theta$ models the secondary control \cite{Weitenberg2017,tchuisseu2018curing}. If the system deviates from the nominal frequency, e.g. because $\omega>0$ for a long time, then the bulk angle $\theta$ increases more and more and thereby the secondary control increases and acts as an increasing force to return the system towards the nominal frequency. We use the simplest integral control, whereas other secondary control implementations \cite{ Weitenberg2017,wang2016control,Simpson-Porco2012,hammid2016load,Dongmo2017,martyr2019benchmarking} might be considered in the future. Typically, the magnitude of the primary control parameter is much larger than the secondary control parameter $c_1 \gg c_2$ to implement that primary control acts faster than secondary control.

The power mismatch is given as $\Delta P$. It contains only the deterministic mismatch between supply and demand. If generation surpasses consumption, $\Delta P$ becomes positive and vice versa.  In our market model, we will employ a time-dependent $\Delta P$, inspired by empirical power trajectories, see Fig.~\ref{fig:DemandVsGeneration}.

Finally, $\epsilon \xi$ denotes the aggregated noise acting on the system. As pointed out in the previous section, we assume $\xi$ to be white Gaussian noise, i.e., its time average is zero $\langle \xi(t) \rangle=0$ and its correlation is zero for non-identical times, i.e., it is a delta function $\langle \xi(t)\xi(t') \rangle=\delta(t-t')$ \cite{Gardiner1985}. Extensions using correlated or non-Gaussian noise are also possible in the same framework.

The model \eqref{eq:model} is very general as we have not yet specified the parameters $c_1$, $c_2$, $\epsilon$ or the function $\Delta P$. Note again the different roles of primary and secondary control: Assume $\Delta P =P_0 > 0$ for a long time, this will increase the angular velocity $\omega$ and thereby the angle $\theta$. Without secondary control and noise, i.e., $c_2= \epsilon =0$, the new quasi--steady state becomes $\omega ^* \approx P_0/c_1>0$. The full fixed point $\omega=0$ can only be restored with an additional (integrative) secondary control.

\subsection{Cases}

\begin{figure*}
\centering
\includegraphics[width=0.8\textwidth]{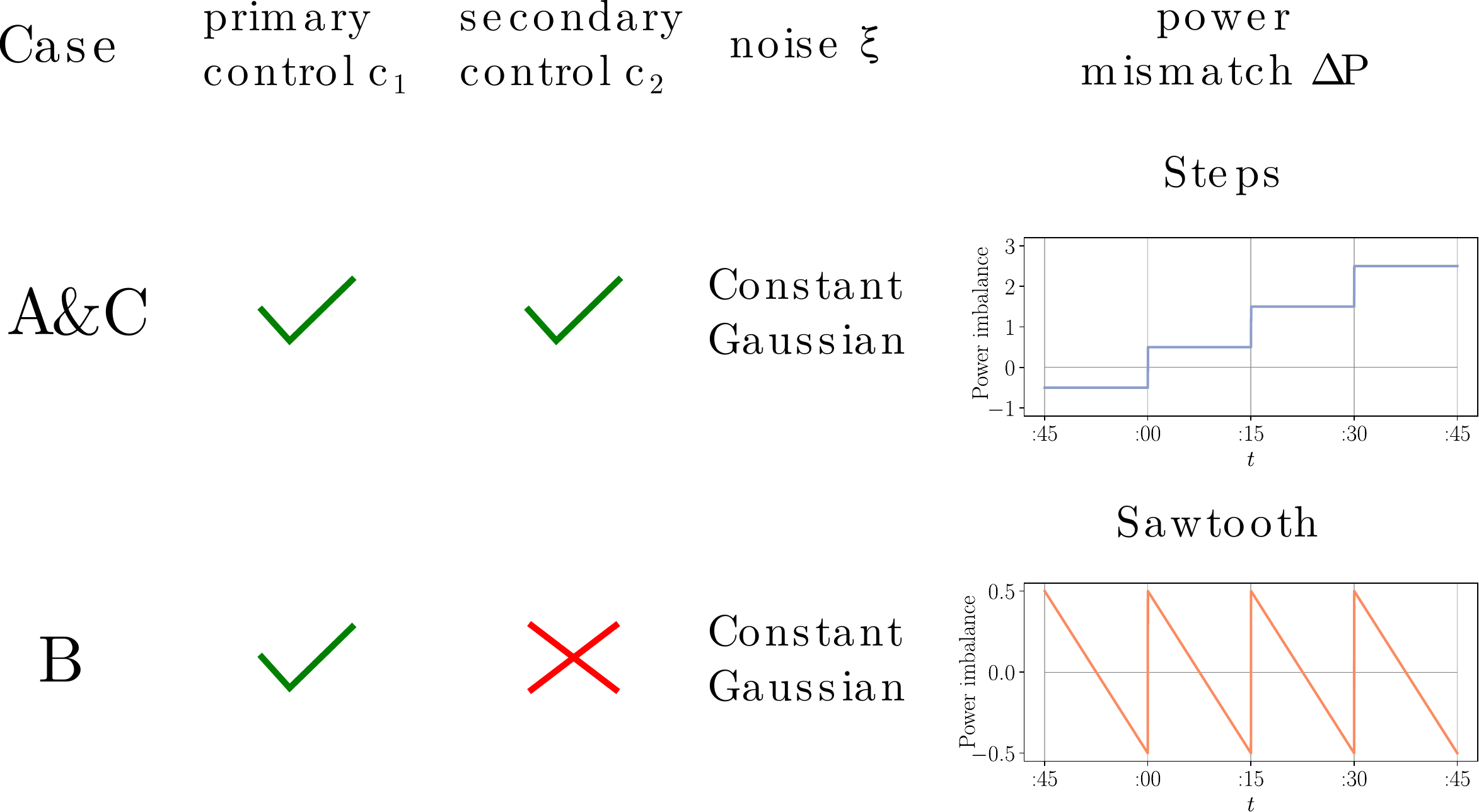}
\caption{{\bf We consider three different cases to model the power-grid frequency based on the model~\eqref{eq:model}.}
In Case A, we apply constant primary and secondary control, white Gaussian noise with constant amplitude $\epsilon$ and a Heaviside power mismatch $\Delta P$.
In contrast, Case B uses no secondary control and applies a sawtooth function for the power mismatch $\Delta P$.
We still apply a constant primary control $c_1$ and white Gaussian noise with constant amplitude $\epsilon$.
Finally, Case C uses Case A's settings but we extract the jump heights of the Heaviside function from independent historic demand data \cite{ENTSOE_Demand} and not from the frequency trajectory.
As in Fig.~\ref{fig:DemandVsGeneration}, we display all jumps with the same height for simplicity.
%Finally, case C again uses constant primary and secondary control, but allows time-varying noise, such as different fluctuations in the morning and evening and the power mismatch is based on realistic demand values.
}
\label{fig:ModelOverview}
\end{figure*}

We consider some special cases of parameter choices for model \eqref{eq:model} here. Theoretically, the model proposed so far would allow that the three parameters $c_1$, $c_2$, and $\epsilon$ are chosen as  zero or non-zero constants, time-dependent functions, or to follow their own stochastic process. Similarly, the power mismatch $\Delta P$ could be any function, as long as the differential equation is still well-defined. We review three cases, see also Fig.~\ref{fig:ModelOverview} for an overview.

The distinguishing factor between those cases is the role of secondary control $c_2$ and power imbalance $\Delta P$: Any non-zero power imbalance $\Delta P$ will be compensated by secondary control if $c_2>0$. This means from a data-analysis it is virtually impossible to distinguish cases where $\Delta P=0$ and no secondary control is active or $\Delta P \neq 0$ and secondary control restored the frequency or a case where a slowly changing $\Delta P$ restored the frequency on its own without secondary control active.
Complementary, large and rapid changes in the power imbalance are clearly visible in the frequency trajectory and always have to be included in the models.

\emph{Case A}: A simple starting point is to set $c_1$, $c_2$, and $\epsilon$ all as non-zero constants.
By including an active secondary control, we neglect slow changes in the power imbalance $\Delta P$ and assume that secondary control is the main restoring force following a sudden jump. Specifically, we assume that the power mismatch $\Delta P$ is given as a piece-wise constant function, i.e., a Heaviside function.
This model has the advantage that we can easily estimate all parameters from the trajectory.
%and use a piece-wise constant power mismatch $\Delta P$, i.e., a Heaviside function. This model has the advantage that we can easily estimate all parameters from the trajectory.

\emph{Case B:} Alternatively, we may neglect the effects of secondary control, setting $c_2=0$.
To balance the frequency, we then require a balanced power dispatch on average, i.e., $\langle \Delta P \rangle=0$.
A simple function to realise this, while maintaining the jumps, which are visible form the frequency trajectories, is a sawtooth function, i.e., piecewise linearly increasing or decreasing over time.
Similar to Case A, we still use constant non-zero $c_1$ and $\epsilon$.

\emph{Case C}: We again repeat Case A but instead of estimating the power mismatch $\Delta P$ from frequency trajectories, we use historic demand data of Germany, based on data published by ENTSO-E \cite{ENTSOE_Demand}.
%\emph{Case C:} 
%To demonstrate how complicated the model can become, we briefly review the possibility of a more realistic power mismatch $\Delta P$, based on data, include a time-dependent noise, while keeping primary and secondary control constant.

%\subsection{Case A: Step function and secondary control}
%\subsection{Case B: Sawtooth function and no secondary control}
%\subsection{Case C: Realistic power mismatch and noise}

\subsection{Estimating parameters}

To generate a synthetic trajectory approximating real data, we need to estimate suitable parameters for our model. 
Here, we present the mathematical background and basics that allow this parameter estimation as well as illustrations of the procedure in Figs.~\ref{fig:EstimatePrimaryControl}, \ref{fig:EstimateEpsilon} and \ref{fig:EstimatePandC2}.
We provide additional guidance and code on how the estimators can be applied in practice in the Supplemental Material.

We estimate the parameters of the synthetic model as follows: %The noise $\epsilon$ is obtained from using the second Kramers--Moyal coefficient. The primary control $c_1$ is given by initially detrending the trajectory and them computing the first  Kramers--Moyal coefficient.
The primary control $c_1$ and the noise $\epsilon$ are obtained from using the first and second Kramers--Moyal coefficient respectively.
Next, the power mismatch $\Delta P$ and the secondary control $c_2$ are determined from the trajectory at the trading times. 

%We  split the analysis in two parts: We initially focus on time intervals before the trading, where the system had time to relax to a stochastic steady state and estimate the noise $\epsilon$ and the primary control $c_1$. Complementary, the trading jumps itself and the decay  back towards the nominal frequency are used to recover $\Delta P$ and $c_2$. 

\subsubsection{Kramers--Moyal and Fokker--Planck}
Let us briefly review some relevant stochastic theory necessary to estimate the parameters.
The synthetic model \eqref{eq:model} includes stochastic and deterministic dynamics.
Assuming that the deterministic contribution given by $\Delta P$ and the secondary control $c_2$ are either very small or subtracted from the trajectory, we are left with a purely stochastic process for $\omega$ in the form of a Langevin equation.
Such an equation cannot be solved deterministically, but we may formulate the Fokker--Planck equation of the stochastic system instead \cite{Gardiner1985}:
%\begin{equation}\label{power_derivation}
%\begin{aligned}
%\frac{\partial p}{\partial t} =  - \frac{\partial}{\partial \omega}\left(-c_1 \omega p \right) + \frac{\epsilon^2}{2} \frac{\partial^2 p}{\partial \omega^2}.
%\end{aligned}
%\end{equation}

\begin{equation}\label{power_derivation}
\begin{aligned}
\frac{\partial p}{\partial t} =&  - \frac{\partial}{\partial \omega}\left(-c_1 \omega p \right) + \frac{\epsilon^2}{2} \frac{\partial^2 p}{\partial \omega^2} \\
& - \frac{\partial}{\partial \omega}\mathcal{D}^{(1)} p  +  \frac{\partial^2 }{\partial \omega^2} \mathcal{D}^{(2)}p.
\end{aligned}
\end{equation}

This Fokker--Planck equation is a partial differential equation for the probability density function $p(\omega,t)$ of the system.
Solving this Fokker--Planck equation thereby returns the probability $p(\omega,t)$ to observe the system in state $\omega$ at time $t$ \cite{Gardiner1985}.

Terms subject to first derivatives are known as \emph{drift terms} $\mathcal{D}^{(1)}$, while terms subject to second derivatives are called \emph{diffusion terms} $\mathcal{D}^{(2)}$ \cite{Gardiner1985}.
Drift terms describe the deterministic behaviour of the full stochastic system, e.g. the movement of a particle within a potential or in our case the control and damping forces acting within the power grid, causes a ``drift'' towards the stable state.
Complementary, the diffusion terms determine the stochastic part of the trajectory.
Random noise makes state of the grid ``diffuse'' through the available state space and typically leads to a broadening of the probability distribution $p$ \cite{Gardiner1985}.
We can read off the drift and the diffusion terms of the angular velocity $\omega$ as  $\mathcal{D}^{(1)}=-c_1 \omega$ and  $\mathcal{D}^{(2)}=\frac{\epsilon^2}{2}$ respectively.
These drift and diffusion terms of the Fokker--Planck equation are also known as the Kramers--Moyal coefficients from the Kramers--Moyal expansion of the fundamental master equation of the system. Only this approximation allows us to write the Fokker--Planck equation \cite{Risken1984a,friedrich2011}. From these coefficients we estimate the mentioned parameters. 

\subsubsection{Estimating the primary control $c_1$}
Having set out the theory of Fokker--Planck equations and Kramers--Moyal coefficients, we now apply them to determine the primary control $c_1$, by applying a two-step process: We first subtract the deterministic and slow time scale components from the trajectory and then determine the first Kramers--Moyal coefficient.

We first remove the driving deterministic characteristics of the model \eqref{eq:model} from any trajectory we analyse.
To do so, we filter the data with a Gaussian kernel filtering, with a window of $60$ seconds, to remove the deterministic trend and any slow process, such as secondary control, and thus remain solely with the stochastic component of the process.
The procedure is independent of the specific driving method (cf. Case A and Case B).

%Figures illustrating the parameter estimation
\begin{figure}
\includegraphics[width=0.99\linewidth]{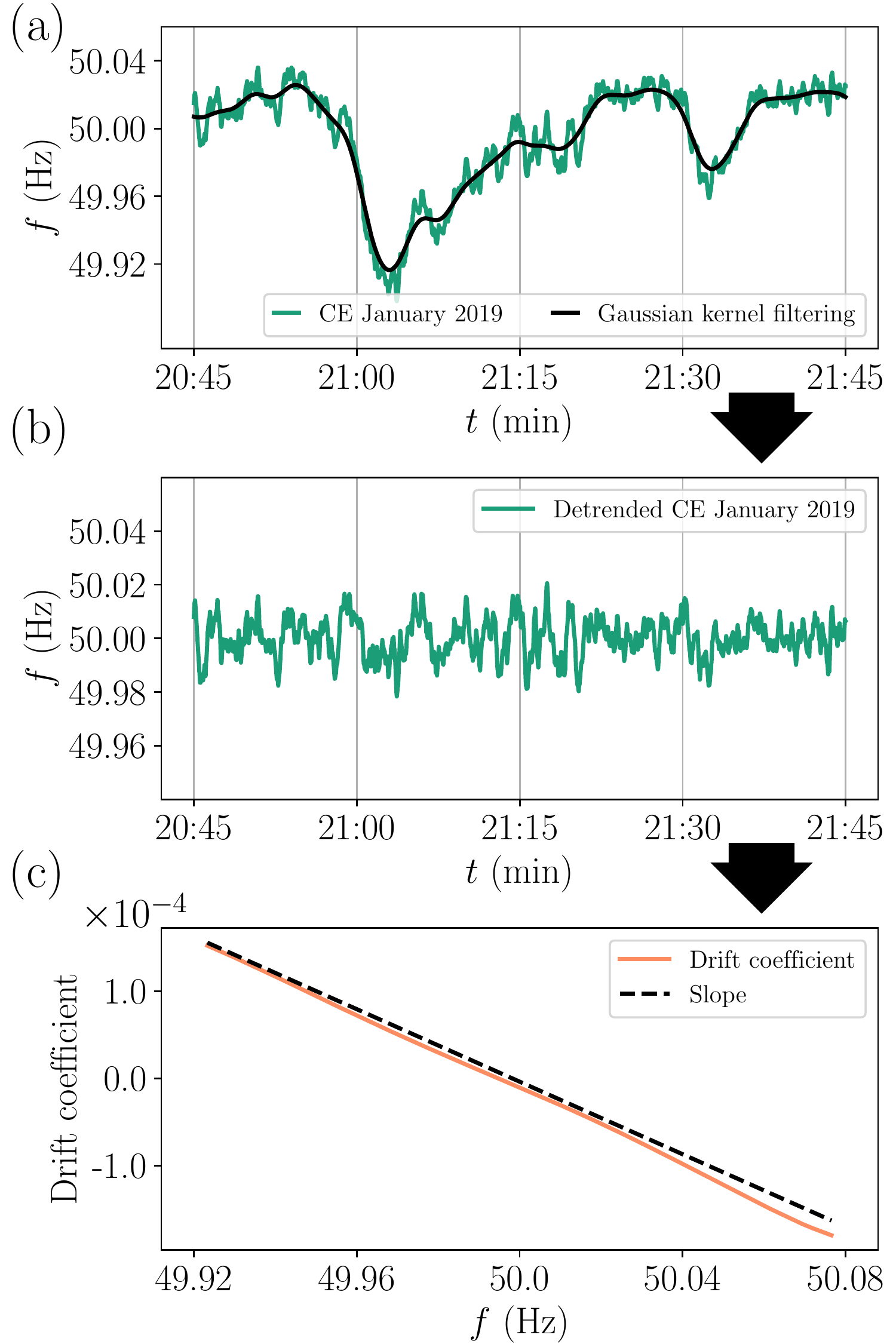}
\caption{{\bf The primary control $c_1$ is computed from fluctuations around the trend}. To estimate the primary control $c_1$, we first detrend the data by applying a Gaussian kernel and then compute the drift coefficient.
{\bf (a)}: We display a snippet of the power-grid frequency trajectory from the CE data from January 2019, as in Fig.~\ref{fig:Trajectory}, alongside with the $60$ seconds window Gaussian kernel detrending, that captures the deterministic and slowly changing contributions of the power-grid frequency.
{\bf (b)}: We extract the stochastic motion by subtracting  the deterministic trend from the power-grid frequency.
%Show the removal of the deterministic trend by subtracting from the power-grid frequency trajectory the Gaussian kernel detrending.
What is left is a stochastic trajectory resembling approximately an Ornstein--Uhlenbeck process.
{\bf (c)}: We compute the first Kramers--Moyal coefficient, known as the drift coefficient, of the now purely stochastic process.
The slope of the drift coefficient is equal to the primary control $-c_1$.}\label{fig:EstimatePrimaryControl}
\end{figure}

The detrending is illustrated in Fig.~\ref{fig:EstimatePrimaryControl}: The same snippet of data from Fig.~\ref{fig:Trajectory} is shown alongside with the Gaussian kernel detrending. In panel (b) the subtraction of the detrending on the data yields the purely stochastic process governing the power-grid frequency dynamics without deterministic or slow time scale influences.
Finally, we extract the first Kramers--Moyal coefficient in panel (c):
\begin{equation}\label{eq:Drift_Coefficient}
    \mathcal{D}^{(1)}(\omega) = \frac{1}{\Delta t}\langle ( \omega ( t + \Delta t ) -  \omega ( t )) |_{\omega(t) = \omega}\rangle = -c_1 \omega,
\end{equation}
where $\Delta t$ is the sampling rate of the process at hand, which is  $\Delta t=1$~s for our data sets. Furthermore, $\langle ...|_{\omega(t) = \omega} \rangle$ denotes the following:
%the time average and  means that we evaluate the difference for point $\omega(t)=\omega$: 
A spatial average of the difference $(\omega(t+\Delta t) -\omega(t))$ is taken at the point of evaluation $\omega(t)=\omega$, i.e., at a particular frequency $\omega$ all differences $(\omega(t+\Delta t) -\omega(t))$ are evaluated and the diffusion $\mathcal{D}^{(1)}$ is obtained as a function of $\omega$. Based on our modelling assumptions, we presume this function to be linear in $\omega$. And, when we apply this to the real data in Fig.~\ref{fig:EstimatePrimaryControl}, we notice that the numerically extracted drift term is indeed well described as a linear function with slope $-c_1$.

\subsubsection{Estimating the noise amplitude $\epsilon$}
%We still assume that deterministic and long-term effects are filtered out of the trajectory and determine the amplitude of the stochastic motion, denoted as $\epsilon$. 
The noise amplitude $\epsilon$ is unravelled from data by studying the second Kramers--Moyal coefficient. In our case, we obtain the second conditional moment as
\begin{equation}\label{eq:Diffusion_Coefficient}
    \mathcal{D}^{(2)}(\omega) = \frac{1}{\Delta t}\langle ( \omega ( t + \Delta t ) - \omega ( t ))^2 |_{\omega(t) = \omega}\rangle = \frac{\epsilon^2}{2},
\end{equation}
where $\Delta t$ is again the sampling rate of the process and the empirical $\mathcal{D}^{(2)}(\omega)$ is assumed to approximately constant, based on our model. Computing the second conditional moment $\mathcal{D}^{(2)}$ thereby yields the noise amplitude $\epsilon$.
Empirically, we note that the de-trending is not even necessary to determine the correct diffusion coefficient. So we instead compute the diffusion from the original data directly.

\begin{figure}
\includegraphics[width=0.99\linewidth]{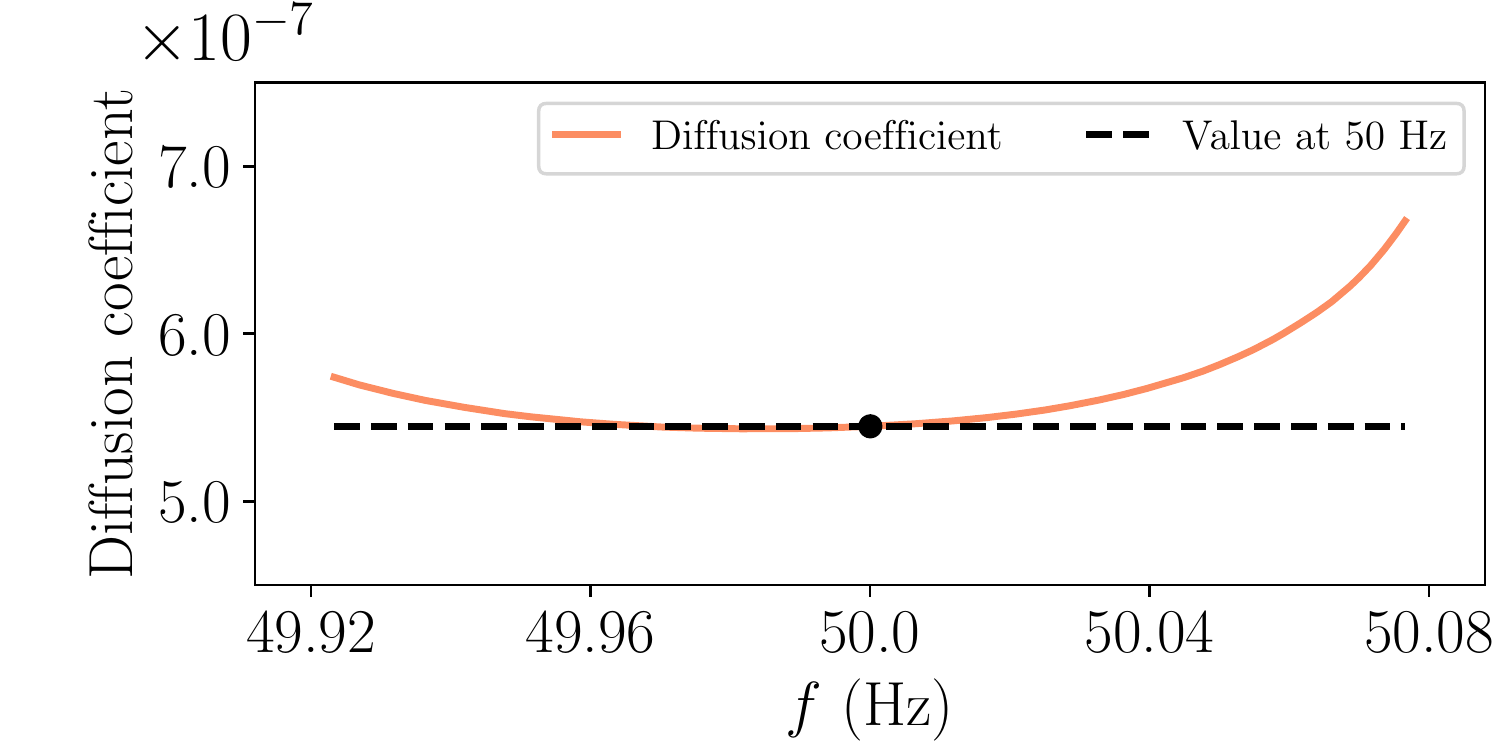}
\caption{{\bf The noise amplitude $\epsilon$ is obtained using the diffusion coefficient}. We display the diffusion coefficient, or second Kramers--Moyal coefficient around $50$~Hz for the CE grid for the month of January 2019.
By taking the value at $50$~Hz, indicated on the plot, and by using relation \eqref{eq:Diffusion_Coefficient}, we obtain the noise $\epsilon$.}\label{fig:EstimateEpsilon}
\end{figure}

We display the diffusion coefficient, i.e., the second Kramers--Moyal coefficient, as a function of the frequency in Fig.~\ref{fig:EstimateEpsilon} for the month of January 2019 for the CE grid. We determine the diffusion coefficient value at $50$~\unit{Hz}  and by using \eqref{eq:Diffusion_Coefficient} thus determine the noise amplitude $\epsilon$.

\subsubsection{Estimating the market impact $\Delta P$}
To determine both $\Delta P$ and $c_2$, we have a closer look at the frequency behaviour following a sudden power imbalance. 
Assuming that the power imbalance is large enough, we can neglect the noise amplitude $\epsilon \approx 0$ as the dynamics close to the power jump are approximately deterministic.
Before the power imbalance, we assume that the system is close to the nominal frequency, i.e., $\Delta P=0$, $\theta \approx 0$ and $\omega\approx 0$.
Next, we introduce a power imbalance, e.g. due to trading by setting $\Delta P=P_0$.
The equations of motion then are
\begin{equation}
    \begin{aligned}\label{eq:DeltaPEquationOfMotion}
    \frac{\d \theta}{\d t}  & = \omega,\\
    \frac{\d \omega}{\d t} & = -c_{1}\omega-c_{2}\theta+ P_0.
    \end{aligned}
\end{equation}
A full solution of this driven, damped harmonic oscillator is given by
\begin{equation}
    \omega(t)=\frac{P_0e^{-\frac{1}{2} t \left(\sqrt{c_1^2-4 c_2}+c_1\right)}}{\sqrt{c_1^2-4 c_2}}\left[e^{t \sqrt{c_1^2-4 c_2}} -1 \right].
    \label{eq:OmegaSolution_afterJump}
\end{equation}

We evaluate the rate of change of frequency (ROCOF) at the jump time, i.e., at $t=0$ to be
\begin{equation}
\left.\frac{\mathrm{d}\omega}{\mathrm{d}t}\right|_{t=0}=P_0,
\end{equation}
and thereby determine the jump height $P_0$, which gives us the power imbalance $\Delta P$, again assuming $\theta(0) =\omega (0)\approx 0$. Recall that we rescaled all variables with the inertia $M$ so that the ROCOF depends on the change of power and the inertia as expected.

Note, while the solution \eqref{eq:OmegaSolution_afterJump} explicitly used the Heaviside function with secondary control (Case A), the ROCOF also determines the power jump in the case of a sawtooth function (Case B). The reason is that the derivative at $t=0$ is independent of what happens for $t>0$ and also does not depend on $c_1$ or $c_2$.

\begin{figure}
\centering
\includegraphics[width=0.98\linewidth]{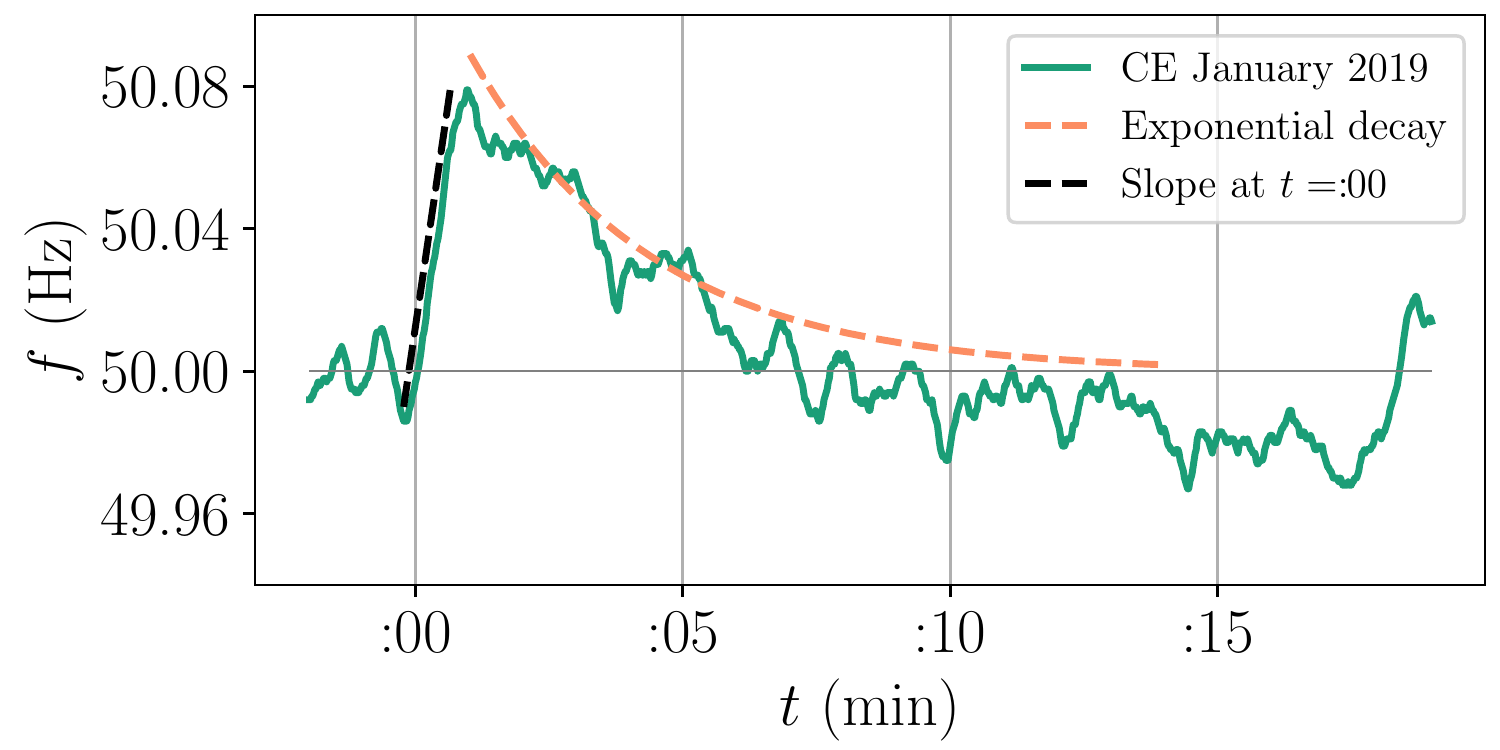}
\caption{{\bf Power imbalance $\Delta P$ and secondary control $c_2$ are determined from trading peaks.}
We investigate the frequency trajectory at a trading peak: The power imbalance $\Delta P$ is obtained from the initial slope, i.e., the rate of change of frequency (ROCOF) and the secondary control $c_2$ from the following exponential decay, see \eqref{eq:OmegaSolution_afterJump}. The frequency trajectory is using the CE data from January 10 2019.}\label{fig:EstimatePandC2}
\end{figure}

\subsubsection{Estimating the secondary control $c_2$}
The estimation of $c_2$ is only necessary for models that include it, such as Case A with its simple Heaviside function.
We know how the trajectory of the angular velocity $\omega$, given by  \eqref{eq:OmegaSolution_afterJump}, develops following a jump:
Initially, the value of $\omega$ increases and then decays approximately exponentially back to the reference value. 

Since the primary control parameter $c_1$ is typically much larger than the secondary control parameter $c_2$, we make use of the following approximation: $\sqrt{c_1^2-4c_2}\approx c_1 -\frac{2 c_2}{c_1}$. Thus \eqref{eq:OmegaSolution_afterJump} reduces to

\begin{equation}\label{eq:approx_of_7}
\omega(t)=\frac{P_0e^{-t \frac{c_2}{c_1}}}{c_1 - \frac{2c_2}{c_1}}\left[1-e^{-t \left(c_1 -\frac{2c_2}{c_1}\right)} \right].
\end{equation}

For larger times $t\gg 1 s$, the second term in \eqref{eq:approx_of_7} decays much faster than the first term. 
We can therefore further approximate the angular velocity $\omega$
as
\begin{eqnarray}\label{eq:exp-decay-omega}
    \omega (t) & \sim & \exp\left(-\frac{c_2}{c_1} t\right),
\end{eqnarray}
which allows an estimate of the secondary control $c_2$, taken we determined the primary control $c_1$ earlier. We only need to determine the exponent of the exponential decay, as depicted in Fig.~\ref{fig:EstimatePandC2}.
Note that the exponential decay constant does not depend on which trading interval we analyse. For more robust analysis, we perform the fits using the decay following hourly jumps, see also Supplemental Material.
%@Benjamin This makes sense, regardless of the RoCoF or NADIR/Overshoot, the exponents of the exponential decay are the same. The data is (of course) better for big jumps, since the decay is clear What should I delete?}

This sequence of parameter estimations allows us to uncover all underlying parameters of the system directly from power-grid frequency measurements.
In fact, a single measurement of $60$ minutes of data already entails a good ground for estimation, but naturally employing as much data as possible yields more reliable parameter estimations, as well as the possibility of error estimation in an efficient way.

\section{Case study: Continental European grid}
With the model properly defined, we now show how it approximates the stochastic behaviour of real frequency trajectories in Europe. 
The frequency statistics and also market setting differ substantially between different power grids \cite{Schaefer2017a}. So, instead of applying each case to all potential power grids, we showcase it on one power grid example where the statistics are well approximated.

Hence, we first apply Case A to data from Continental Europe, Case B to data from Great Britain and finally show that we can also import and utilize real dispatch data to further improve the model predictions in Case C.

\begin{figure}[h!]
\centering
\includegraphics[width=0.99\linewidth]{{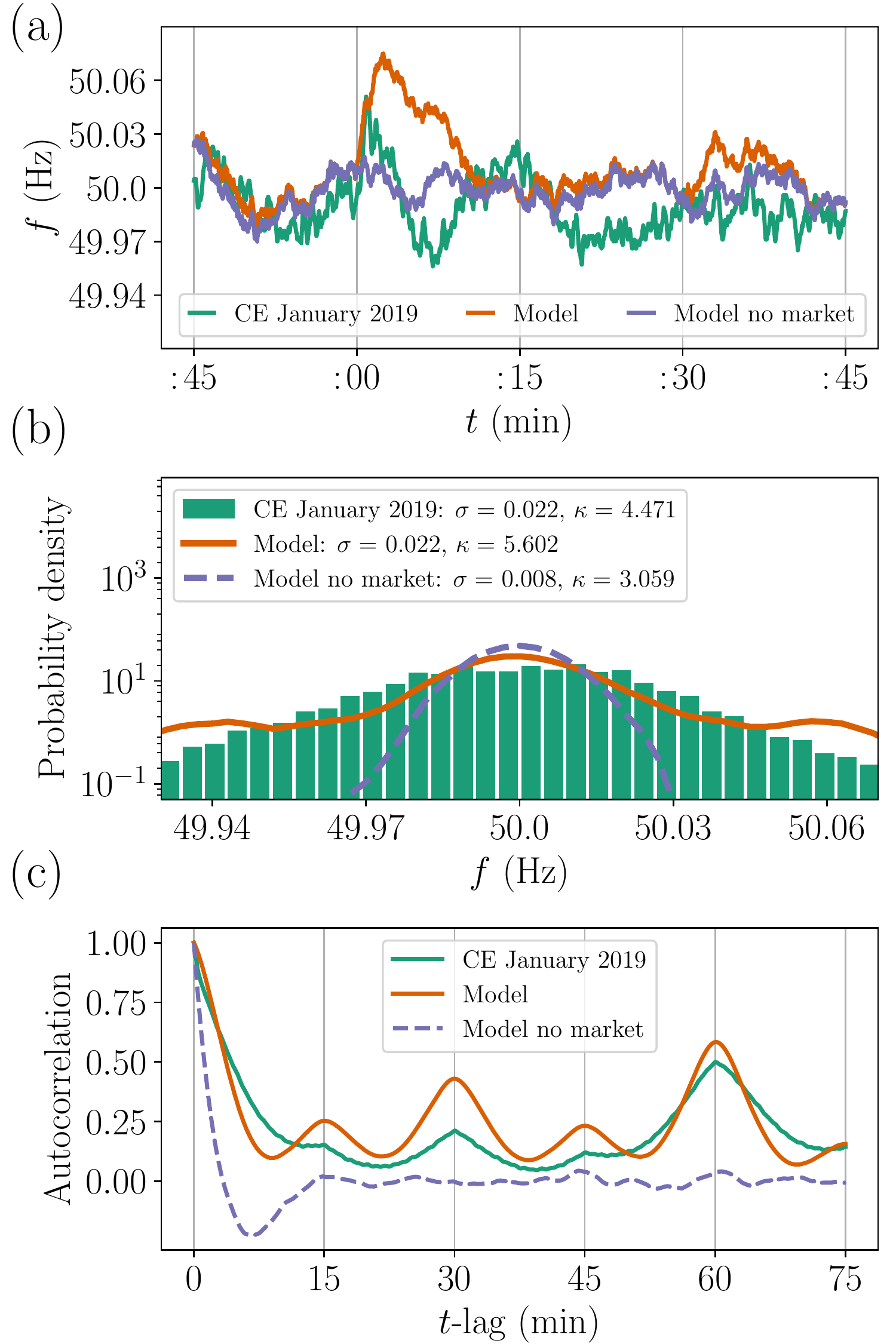}}
\caption{{\bf Case A: Heaviside dispatch approximates CE trajectories}. 
We compare two days of the power-grid frequency of the Central European (CE) power grid in January 2019 with synthetic data generated by our model \eqref{eq:model}.
For this particular analysis, we utilise Case A that relies on a step function mimicking the jumps of the  power mismatch $\Delta P$.
 %Sec.~\ref{Sec:guidelines}, 
The four governing parameters: Noise $\epsilon$, primary $c_1$ and secondary $c_2$ control, and power mismatch $\Delta P$ parameters are given in Section \ref{sec:caseA},  further details are given in the Supplemental Material. 
{\bf (a)} We plot a snippet of the power-grid frequency trajectory from the CE data, the model, and the surrogate model without power dispatch.
The 15 minute trading intrinsic to the model and the data is highlighted with grey lines.
{\bf (b)} We display the probability density function of the CE data (histogram), the model data (solid line), and the surrogate model data without dispatch (dashed line).
Standard deviation and kurtosis of each process are indicated in the legend.
{\bf (c)} We display the autocorrelation of the processes for a time window of 75 minutes, noting the initial exponential decay and regular peaks.}\label{fig:CE_Jan2019}
\end{figure}

\subsection{Case A: Parameter extraction for January 2019, Central Europe}
\label{sec:caseA}
We analyse power-grid frequency data for the month of January 2019, using measurements provided by the  transmission system operator TransnetBW GmbH who operates the German grid in the state Baden-W\"urttemberg \cite{Transnet}. For this month, we estimate the following parameters:

\begin{center}
 \begin{tabular}{c c c c} 
 \multicolumn{3}{c}{\textbf{Central Europe}, January, 2019. }\\
 \hline\hline
 $\epsilon~[s^{-2}]$\quad~ & $c_1~[s^{-1}]$ & $c_2~[s^{-2}]$ \\ [0.5ex] 
 \hline
 $0.00105$ & $0.008311$ & $0.000030$ \\ 
\end{tabular}
\end{center}
%$P_0~[s^{-2}]$ & $0.001641$ & 

For simplicity, we considered the dispatch at the hourly mark as the reference, as can be seen in Fig.~\ref{fig:Trajectory} to be the strongest driver of the system.

\begin{center}
 \begin{tabular}{c | c c c} 
 \multicolumn{4}{c}{\textbf{Case A:} CE $P_0$, January, 2019}\\
 \hline\hline
 ~~~\qquad~ & ~~~at $:\!00$ & \quad at $:\!30$ & at $:\!15$, $:\!45$ \\ [0.5ex] 
 \hline
 $P_0~[s^{-2}]$ & ~~~$0.001641$ & \quad $0.000547$ &  $0.000273$ \\ [0.5ex] 
\end{tabular}
\end{center}

Extracting the value, as described, of the $\Delta P$ for the hourly mark, we considered the half-hour and quarter-hour trading windows to be $1/3$ and $1/6$ of the hourly value of $\Delta P$.
Notice that there is no limitation in calculating this from data, but the results can prove unreliable given the small differences in dispatch.
Furthermore, to mimic the structure of the dispatch \cite{Weisbach2009}, we take a naïve $6$-hour window where the $P_0$ jumps are positive values, followed by an equivalent $6$-hour window with negative peaks.
This should approximate the daily cycles of human daily activity: The work schedule begins: demand increases; Work schedule ends: demand decreases; Private consumption at home begins: demand increases; Night time begins, demand decreases.

\begin{center}
\begin{tabular}{c c c c} 
 \multicolumn{4}{c}{\textbf{Case A:} CE power mismatch pattern}\\
 \hline\hline
 ~$02{:} \!-\! 08{:}$~ & ~$08{:} \!-\! 14{:}$~ & ~$14{:} \!-\! 20{:}$~ & ~$20{:} \!-\! 02{:}$~ \\ [0.5ex] 
 \hline
 $\searrow$ & $\nearrow$ & $\searrow$ & $\nearrow$ \\ [0.5ex] 
\end{tabular}    
\end{center}

Having these parameters at hand, we can now employ our model \eqref{eq:model} to integrate synthetic power-grid frequency trajectories. 
We employ an Euler--Mayurama stochastic integrator, with a time sampling of $0.001$ seconds, for a total length of two days, and make use on a step function with changing values every $15$ minutes, as formulated in Case A, to mimic the power dispatch curve.

We compare the data, the synthetic model based on \eqref{eq:model} and surrogate model without the market structure, i.e., where we set $\Delta P=0$, in Fig.~\ref{fig:CE_Jan2019}.
%In Fig.~\ref{fig:CE_Jan2019} are depicted the main observations of the model \eqref{eq:model}, in comparison both with the data and the equivalent surrogate model \eqref{eq:model} without the market structure.
The introduction of the model without the market allows us to understand concisely the influence of the dispatch on the trajectory of the power-grid frequency, as well as the influence it has on the statistical behaviour of the system.

Several distinct features of the market effect can be seen in Fig.~\ref{fig:CE_Jan2019}:
While the surrogate only fluctuates randomly close to the reference frequency, both the real and the synthetic trajectory display surges of the frequency close to the 15 minute trading windows, see panel (a).
These large surges lead to a non-Gaussian probability distribution of the power-grid frequency, evidenced in panel (b).
Both the data and the synthetic model with the market display a high kurtosis ($\kappa>3$), while the surrogate model without any market is essentially Gaussian. This  indicates that the market activity has a considerable impact on the distribution of the frequency, specifically its tails. With the market, the system reaches critical values much more often than what would be expected by a normally distributed process.
We finally compare the  autocorrelation functions of the power-grid frequency for the CE data of January 2019, the modelled data, and the surrogate model in panel (c).
We note that the system's scheduled trading/dispatch windows generate defined peaks at exactly $15$, $30$, $45$, and $60$ minutes.
By comparison, a surrogate system without a market structure displays no correlation peaks at any time lag.
Moreover, it is importance to notice that \textit{all} peaks in the autocorrelation function are positive valued, both for the synthetic and the real data.
This indicates that the system's dispatch is not an uncorrelated random process but the direction of the frequency change is correlated: Frequency surges are more likely followed by more frequency surges and vice versa for frequency sags.
The modelled data mimics this with accuracy by implementing an over-simplistic yet successful heuristic argument based on human daily cycles, as explained before.

\begin{figure}[h!]
\includegraphics[width=0.99\linewidth]{{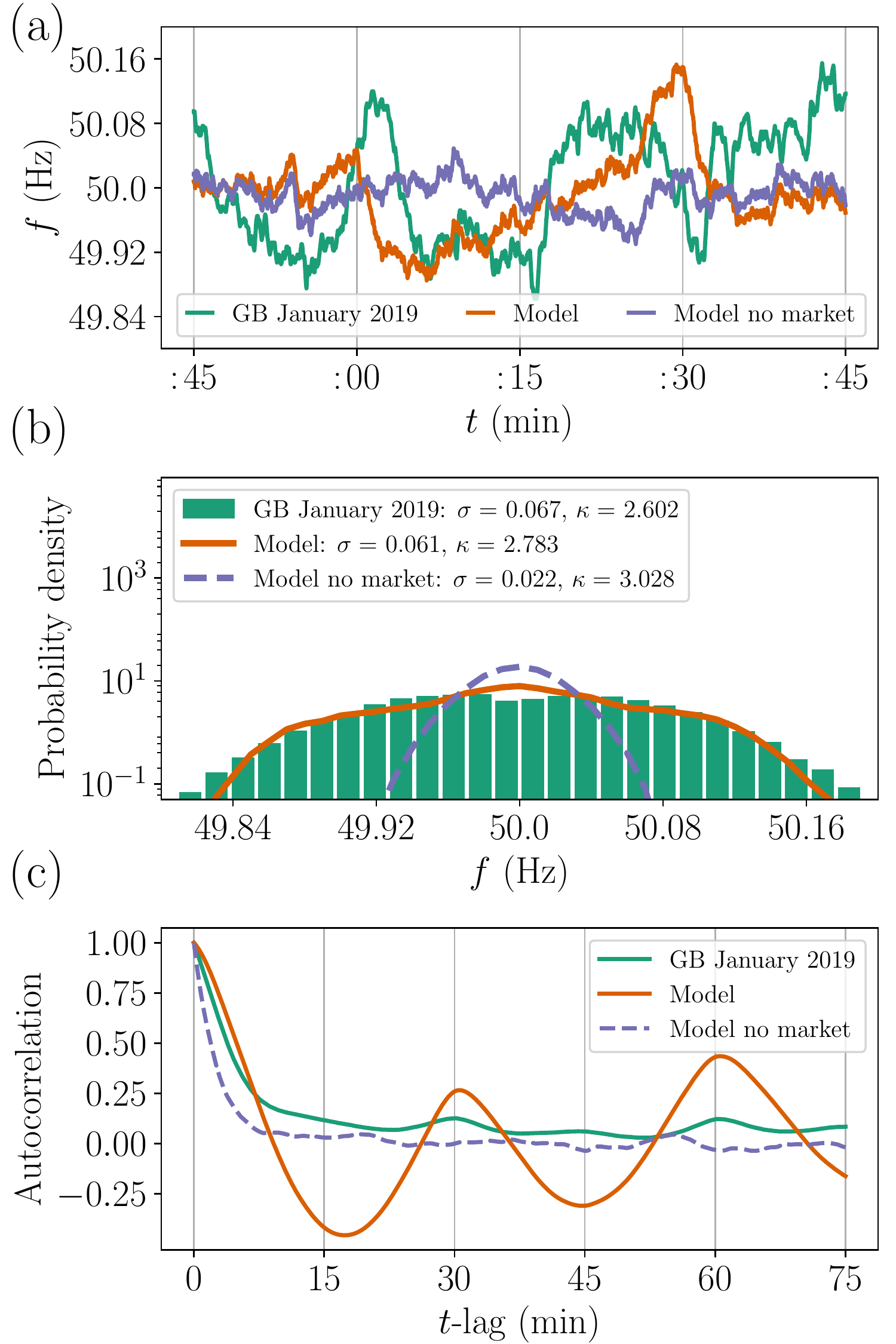}}
\caption{{\bf Case B: Sawtooth dispatch approximates GB trajectories}. We compare two days of the power-grid frequency of the British (GB) power grid in January 2019 with synthetic data generated by our model \eqref{eq:model}.
Here, a sawtooth function is used to describe the mismatch in power $\Delta P$, see Fig.~\ref{fig:ModelOverview}, Case B.
Noise amplitude $\epsilon$, primary $c_1$ control, and power mismatch $\Delta P$ are given in Section \ref{sec:caseB}, see also Supplemental Material for details on parameter estimation. %Sec.~\ref{Sec:guidelines}.
 Note that Case B does not use secondary control.
{\bf (a)} We plot snippet of the power-grid frequency trajectory from the GB data, the model, and the surrogate model without power dispatch.
{\bf (b)}  We display the probability density function of the GB data (histogram), the model data (solid line), and the surrogate model data without dispatch (dashed line).
Standard deviation and kurtosis of each process are indicated in the legend.
{\bf (c)} We display the autocorrelation of the processes for a time window of 75 minutes, noting the initial exponential decay and regular peaks.
Contrary to the Heaviside function of Case A, the sawtooth function forces a negative correlation of the system by first driving the system driven to one  state and then inverting this trend at the trading interval.
%Given the driving implemented in the GB case, from the choice of a sawtooth function, we inherently force a negative correlation of the system, since the system driven to a state and at the trading interval strongly inverted. The tendency to relax along the sawtooth-line curve it antithetically modified my the dispatch jump, leading to negative correlations evidenced in the autocorrelation function.
}\label{fig:GB_Jan2019}
\end{figure}

% In Fig.~\ref{fig:CE_Jan2019} several distinct features of the effects of the market can be seen.
% Notice evidently on panel (a) the surge of the frequency of the model in comparison with the equivalent surrogate with the market.
% This produces a non-Gaussian distribution of the power-grid frequency, evidenced in panel (b).
% The high kurtosis ($\kappa>3$) indicates that the market structure has a considerable impact on the distribution of the frequency, leading to the system's frequency reaching critical values much more often than what would be expected by a normally distributed process.
% Panel (c) exhibits the autocorrelation function of the power-grid frequency for the CE data of January 2019, the modelled data, and the surrogate model.
% One can see that the system's scheduled trading/dispatch windows generate defined peaks at exactly $15$, $30$, $45$, and $60$ minutes.
% By comparison, a system without a market structure displays no correlation at any time lag.
% Moreover, it is importance to notice that \textit{all} peaks in the autocorrelation function are positive valued.
% This indicates that the system's dispatch is not a random process, and there exist correlation between a frequency surge or sag at some point in time, with a future or past point in time.
% The modelled data mimics this with accuracy by implementing a over-simplistic yet successful heuristic argument, as explained before.

\subsection{Case B: Parameter extraction for January 2019, Great Britain}
\label{sec:caseB}

\begin{figure*}[t]
\includegraphics[width=0.99\linewidth]{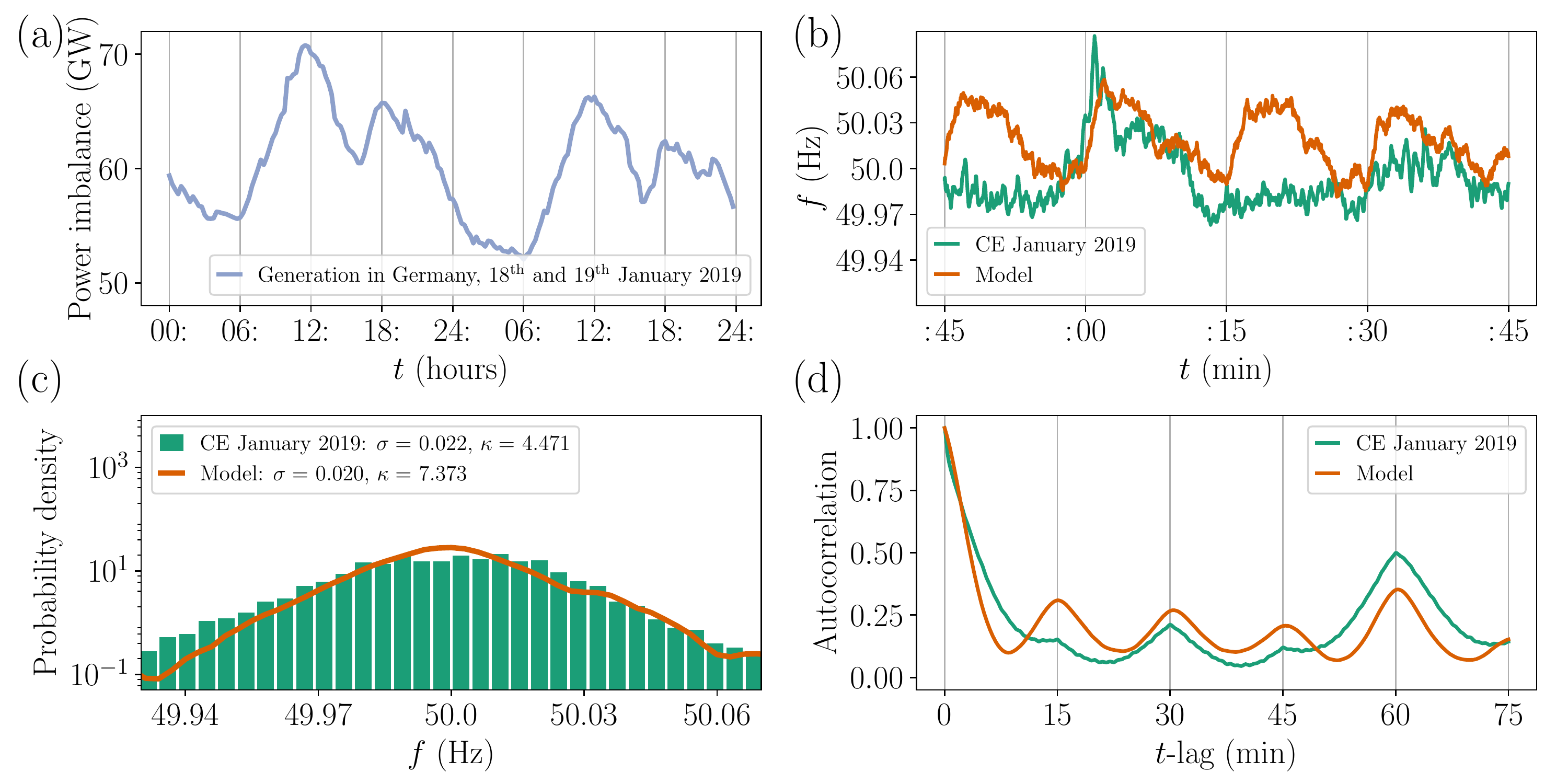}
\caption{{\bf Case C:  Realistic dispatch trajectories better approximate the real frequency statistics}. {\bf (a)} We use the real dispatch trajectories of the demand in Germany \cite{ENTSOE_Demand} to obtain the correct jumps for the Heaviside function (as in Case A) of the power mismatch $\Delta P$  and use our model \eqref{eq:model} to generate a synthetic trajectory. {\bf (b)} The synthetic frequency trajectory statistically resembles the real trajectory for the two day period depicted here. 
Noise amplitude $\epsilon$, primary $c_1$ and secondary $c_2$ control were calculated as described in the Supplemental Material. %Fig.~\ref{fig:CE_Jan2019}. 
{\bf (c)} We display the probability density function of the CE data (histogram), the model data (solid line), and the surrogate model data without dispatch (dashed line).
Standard deviation and kurtosis of each process are indicated in the legend.
{\bf (d)} We display the autocorrelation of the processes for a time window of 75 minutes, noting the initial exponential decay and regular peaks.
While the autocorrelation and the rough shape of the histogram of the model data closely match those of the real data set, we note a substantial difference in the computed kurtosis values. 
This discrepancy is likely caused by the large variations in the volume of the dispatched power. Here we use data from the German grid to allow a 15 minute resolution. However, the full power dispatch affecting the Continental European grid is given as the sum over all participating countries and would likely be smoother and lead to lower kurtosis values. 
%Leo's version: 
%The large variations of the power inbalance (Shouldn't it be called power dispatched? it is not an inbalance at all...) are taken from the German energy dispatch, but the CE system is synchronous in all of Central Europe, thus the dispatch in other European countries contributes similarly to the frequency dynamics.
%The comparison serves to shows that the dispatch structure has a major impact on the power-grid freqency. The mismatch of the kortusis is explained by the difference indicated above.
}\label{fig:RealDispatch}
\end{figure*}

Analogously, we analyse data from Great Britain for the month of January 2019, obtained from the British transmission system operator National Grid ESO \cite{NationalGrid2019}.
Applying the discussed methods, we derive the following parameters
\begin{center}
 \begin{tabular}{c c c c} 
 \multicolumn{4}{c}{\textbf{Great Britain}, January, 2019}\\
 \hline\hline
 $\epsilon~[s^{-2}]$ & $\Delta P~[s^{-2}]$ & $c_1~[s^{-1}]$ & $c_2~[s^{-2}]$ \\ [0.5ex] 
 \hline
 $0.00205$ & $0.00204$ & $0.00606$ & \# \\ 
\end{tabular}
\end{center}
where in this case we set the value of the secondary control $c_2$ to be zero.
%a value of the secondary control $c_2$ can be calculated, but is likely erroneous.
Here, we apply Case B, for two reasons: First, we wish to show that it is also capable of capturing the frequency distributions of a given grid. Second, the British frequency trajectory does not display any clear exponential decay following the trading activity. This is likely caused by a smaller relative trading volume and a larger relative noise amplitude \cite{Schaefer2017a}. Both effects also contribute to much smaller autocorrelation peaks at the trading intervals.
%given the system exhibits frequency trajectories that do not seem to decay exponentially.

We recover the statistics of the British power-grid frequency data with remarkable precision using a sawtooth function for the power dispatch $\Delta P$, see Fig.~\ref{fig:GB_Jan2019}.
The GB data exhibits low kurtosis values ($\kappa<3$), especially when compared to the Continental European values.
Our employed model captures the process with high accuracy, when we apply a sawtooth function for $\Delta P$ (Case B). Notably, for the case of the surrogate model without market activity, the probability distribution again approximates a Gaussian distribution with kurtosis $\kappa=3$.
Although the autocorrelation function exacerbates the peaks, it captures the initial decay and the trend of regular peaks well. The oversized oscillations arise since we assumed consistent periods of six hours with the same jump and ramp behaviour. In turn, the negative autocorrelation arises as the sawtooth function suddenly changes the sign of the market effect. Both assumptions arepart of a very simple but thereby easy-to use model of the British grid, which still captures the probability distribution (histogram) very well.

%Case A is shown for the Continental European grid, displaying a close match of both the histogram and the autocorrelation function. In addition, we apply a sawtooth function in Case B, which again approximates the observed histogram well but exaggerates the autocorrelation function behaviour due to forced sharp up-and down-wards moevemnts.

% As can be seen in Fig.~\ref{fig:GB_Jan2019}, employing Case B, with a sawtooth-like behaviour describing the power-mismatch, recovers the statistics of the British power-grid frequency data with remarkable precision.
% Taking into comparison to the data from Europe, (CE January 2019), the GB data exhibits low kurtosis values ($\kappa<3$).
% The model employed, with the sawtooth-like behaviour, captures the process with accuracy. Notably, without market activity, the distribution again approximates a Gaussian distribution with kurtosis $\kappa=3$.
% Although the autocorelation function exacerbates the peaks, this is due solely to the oversimplification in considering equivalent differences in power demands over the period of four hours.

\subsection{Case C: Using real power dispatch for Continental Europe}

Finally, we use real dispatch data from Germany, provided by ENTSO-E \cite{ENTSOE_Demand}, to determine the power mismatch $\Delta P$  in our model \eqref{eq:model} and compare synthetic and real trajectories in Fig.~\ref{fig:RealDispatch}.
To this end, we simply set the power mismatch $\Delta P$ as a Heaviside function based on the real demand for the German grid, i.e., we use the actual demand and assume it stays constant for a given 15 minute interval. 
As noted before, we only require the jump height in $\Delta P$, here as the demand, while the generation enters as the simplified secondary control term $-c_2 \theta$.
We chose the German data because its time resolution of $\Delta P$ is 15 minutes, compared to 1 hour resolution for many other countries. Using such real demand data breaks the symmetrical and regular six hour patterns we have been using so far in Cases A and B. Thereby, we also include larger time scales in the synthetic frequency data since the real demand naturally includes for example daily and weekly cycles.
Aside form $\Delta P$, we use the same values as in Case A for the other parameters, i.e., noise $\epsilon$, primary and secondary control $c_1$ and $c_2$.
Comparing the synthetic trajectory and derived measures with the real frequency trajectory, we not that including the real demand data improves the approximation  further, see Fig.~\ref{fig:RealDispatch}. For example the probability density of the real frequency is even better approximated by the synthetic data than in Case A.

\section{Discussion}

%Summary: Scope of the paper
We set out to devise a model to generate realistic synthetic trajectories of the power-grid frequency 
to be used in simulations of power and control system dynamics and to assist planning and operation of today's and future power grids.
%for planning of today's and future grids. 
To that end, we first showed that the frequency trajectories show both deterministic and stochastic features, leading to non-standard frequency statistics: Heavy tails in the probability distributions and regular autocorrelation peaks pose challenges to properly model the trajectories.

%Model summary
We proposed a simple model combining the deterministic and stochastic aspects of the trajectories. Using stochastic theory and data analysis we were able to extract all essential parameters of the model from real trajectories. We specifically highlighted how the model approximates probability distributions and autocorrelation functions of realistic grids. A more detailed analysis of the mathematical properties of both real trajectories and the model is presented in \cite{Anvari2019}.

%Model limitations
%The presented model is simple and easy to apply but thereby includes several simplifications: 
The presented model was designed to be generally applicable, easily extendible and usable, which inevitably requires several simplifications:
It does not capture the very short time scale when short-term noise, dynamical behaviour of the rotation machines or switching delays play an important role. Similarly, the model does also not include the long time scale with effects such as synoptic or even seasonal  cycles, long-term trading commitment etc.
Finally, the model is a stochastic model, i.e., it is not suitable for forecasting of the near future but instead it reproduces critical statistical properties such as large frequency deviations.
Conceptually, our modelling approach bridges  power engineering, stochastic modelling and data analysis. Power engineering serves as the inspiration to our model building blocks like primary and secondary control. The universality of stochastic modelling is used in formulating the Fokker--Planck equation and deriving both the diffusion coefficient and primary control. Finally, more data analysis tools are necessary to estimate remaining parameters such as the secondary control or the strength of the dispatch or market actions.

%Market determine heavy tails
Critically, we unveiled how much the market activity influences the tails of the probability distribution, i.e., the probability to observe large deviations from the reference. Comparing models with and without market revealed that just by including the market activity most large events can be explained, consistent with earlier findings \cite{Weissbach2009,Schafer2018b}. This emphasizes the role the market design has on the stability of the power grid.

%Market role in the future
The explicit modelling of the market in the stochastic model is specifically interesting when designing new market rules or introducing new business models. As we have seen, the market has a dramatic influence on the stability-defining large deviations. Our model can easily predict the effects on the frequency when shifting from 15-minute to 5-minute dispatch actions or when introducing real-time pricing. New proposals of smart grids, the impact of demand-side management etc. can all be captured by appropriately modifying the power dispatch $\Delta P$ of our model. Thereby, we provide guidelines how new concepts and devices can be introduced in the grid without destabilizing it but ideally providing additional stability.

%Consequences/Conclusion
Concluding, our research offers a tool that can be used by natural scientists, mathematicians, engineers, economists or industry practitioners on various questions related to the electricity system. It can be used to plan future grids, such as setting up smart grids and microgrids by providing guidelines on how control parameters should be set to guarantee a certain frequency quality. Executable computer code and easy-to-read pseudo-code of the model and the parameter estimation are provided in the supplementary material.

%Outlook
The model presented here can easily be extended in multiple directions: We could apply more advanced stochastic measures to compare the synthetic trajectory with the real trajectory, as partially done in \cite{Anvari2019}. 
Simultaneously, the frequency dynamics considered here could be extended by voltage amplitude dynamics.
Finally, while we only considered constant Gaussian noise, this noise could easily be extended: Either by including explicit non-Gaussian noise \cite{Schaefer2017a}, as it is observed from wind and solar generators \cite{Anvari2016} or by making the noise or the control time-dependent, leading to superstatistical modelling \cite{Beck2003}.

\begin{acknowledgments}
We gratefully acknowledge support from the Federal Ministry of Education and Research (BMBF grant no. 03SF0472 and 03EK3055), the Helmholtz Association (via the joint initiative ``Energy System 2050 - A Contribution of the Research Field Energy'' and the grant No. VH-NG-1025) and the German Science Foundation (DFG) by a grant toward the Cluster of Excellence ``Center for Advancing Electronics Dresden'' (cfaed). This project has received funding from the European Union’s Horizon 2020 research and innovation programme under the Marie Sk\l{}odowska-Curie grant agreement No. 840825.

\includegraphics[width=0.4\textwidth]{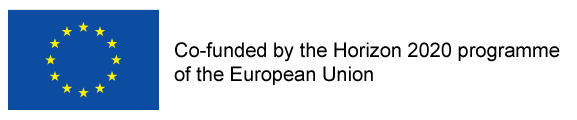}
\end{acknowledgments}

\bibliographystyle{naturemag}
\bibliography{main}

\begin{thebibliography}{10}
\expandafter\ifx\csname url\endcsname\relax
  \def\url#1{\texttt{#1}}\fi
\expandafter\ifx\csname urlprefix\endcsname\relax\def\urlprefix{URL }\fi
\providecommand{\bibinfo}[2]{#2}
\providecommand{\eprint}[2][]{\url{#2}}

\bibitem{sawin2018renewables}
\bibinfo{author}{Sawin, J.~L.} \emph{et~al.}
\newblock \bibinfo{title}{Renewables 2018-global status report}
  (\bibinfo{year}{2018}).

\bibitem{rodriguez2014business}
\bibinfo{author}{Rodr{\'\i}guez-Molina, J.},
  \bibinfo{author}{Mart{\'\i}nez-N{\'u}{\~n}ez, M.},
  \bibinfo{author}{Mart{\'\i}nez, J.-F.} \& \bibinfo{author}{P{\'e}rez-Aguiar,
  W.}
\newblock \bibinfo{title}{Business models in the smart grid: Challenges,
  opportunities and proposals for prosumer profitability}.
\newblock \emph{\bibinfo{journal}{Energies}} \textbf{\bibinfo{volume}{7}},
  \bibinfo{pages}{6142--6171} (\bibinfo{year}{2014}).

\bibitem{Spiegel2019}
\bibinfo{author}{Schultz, S.}
\newblock \bibinfo{title}{{Deutsche Netzbetreiber kämpften mit akuter
  Stromnot}}.
\newblock \bibinfo{howpublished}{{SPIEGEL ONLINE}
  \url{https://www.spiegel.de/wirtschaft/unternehmen/stromnetz-deutsche-netzbetreiber-kaempften-mit-akuter-stromnot-a-1275323.html}}
  (\bibinfo{year}{2019}).

\bibitem{panwar2011role}
\bibinfo{author}{Panwar, N.}, \bibinfo{author}{Kaushik, S.} \&
  \bibinfo{author}{Kothari, S.}
\newblock \bibinfo{title}{Role of renewable energy sources in environmental
  protection: A review}.
\newblock \emph{\bibinfo{journal}{Renewable and Sustainable Energy Reviews}}
  \textbf{\bibinfo{volume}{15}}, \bibinfo{pages}{1513--1524}
  (\bibinfo{year}{2011}).

\bibitem{tuballa2016review}
\bibinfo{author}{Tuballa, M.~L.} \& \bibinfo{author}{Abundo, M.~L.}
\newblock \bibinfo{title}{A review of the development of smart grid
  technologies}.
\newblock \emph{\bibinfo{journal}{Renewable and Sustainable Energy Reviews}}
  \textbf{\bibinfo{volume}{59}}, \bibinfo{pages}{710--725}
  (\bibinfo{year}{2016}).

\bibitem{Obama2013}
\bibinfo{author}{Obama, B.~H.}
\newblock \bibinfo{title}{Presidential policy directive 21: Critical
  infrastructure security and resilience}.
\newblock \emph{\bibinfo{journal}{Washington, DC}}  (\bibinfo{year}{2013}).

\bibitem{Kundur1994}
\bibinfo{author}{Kundur, P.}, \bibinfo{author}{Balu, N.~J.} \&
  \bibinfo{author}{Lauby, M.~G.}
\newblock \emph{\bibinfo{title}{Power system stability and control}},
  vol.~\bibinfo{volume}{7} (\bibinfo{publisher}{McGraw-hill, New York},
  \bibinfo{year}{1994}).

\bibitem{Anvari2016}
\bibinfo{author}{Anvari, M.} \emph{et~al.}
\newblock \bibinfo{title}{Short term fluctuations of wind and solar power
  systems}.
\newblock \emph{\bibinfo{journal}{New Journal of Physics}}
  \textbf{\bibinfo{volume}{18}}, \bibinfo{pages}{063027}
  (\bibinfo{year}{2016}).

\bibitem{wolff2019}
\bibinfo{author}{Wolff, M.~F.} \emph{et~al.}
\newblock \bibinfo{title}{Heterogeneities in electricity grids strongly enhance
  non-gaussian features of frequency fluctuations under stochastic power
  input}.
\newblock \emph{\bibinfo{journal}{arXiv preprint arXiv:1908.07997}}
  (\bibinfo{year}{2019}).

\bibitem{Rohden2012}
\bibinfo{author}{Rohden, M.}, \bibinfo{author}{Sorge, A.},
  \bibinfo{author}{Timme, M.} \& \bibinfo{author}{Witthaut, D.}
\newblock \bibinfo{title}{{Self-organized Synchronization in Decentralized
  Power Grids}}.
\newblock \emph{\bibinfo{journal}{{Physical Review Letters}}}
  \textbf{\bibinfo{volume}{109}}, \bibinfo{pages}{064101}
  (\bibinfo{year}{2012}).

\bibitem{Walter2014}
\bibinfo{author}{Walter, T.}
\newblock \bibinfo{title}{{Smart Grid neu gedacht: Ein L\"osungsvorschlag zur
  Diskussion in VDE|ETG}} (\bibinfo{year}{2014}).
\newblock
  \urlprefix\url{http://www.vde.com/de/fg/ETG/Pbl/MI/2014-01/Seiten/Homepage.aspx}.

\bibitem{Schaefer2015}
\bibinfo{author}{Sch{\"a}fer, B.}, \bibinfo{author}{Matthiae, M.},
  \bibinfo{author}{Timme, M.} \& \bibinfo{author}{Witthaut, D.}
\newblock \bibinfo{title}{{Decentral Smart Grid Control}}.
\newblock \emph{\bibinfo{journal}{{New Journal of Physics}}}
  \textbf{\bibinfo{volume}{17}}, \bibinfo{pages}{015002}
  (\bibinfo{year}{2015}).

\bibitem{Fang2012}
\bibinfo{author}{Fang, X.}, \bibinfo{author}{Misra, S.}, \bibinfo{author}{Xue,
  G.} \& \bibinfo{author}{Yang, D.}
\newblock \bibinfo{title}{{Smart Grids - The new and improved Power Grid: A
  Survey}}.
\newblock \emph{\bibinfo{journal}{{Communications Surveys \& Tutorials, IEEE}}}
  \textbf{\bibinfo{volume}{14}}, \bibinfo{pages}{944--980}
  (\bibinfo{year}{2012}).

\bibitem{weber2018wind}
\bibinfo{author}{Weber, J.} \emph{et~al.}
\newblock \bibinfo{title}{Wind power persistence is governed by
  superstatistic}.
\newblock \emph{\bibinfo{journal}{arXiv preprint arXiv:1810.06391}}
  (\bibinfo{year}{2018}).

\bibitem{Filatrella2008}
\bibinfo{author}{Filatrella, G.}, \bibinfo{author}{Nielsen, A.~H.} \&
  \bibinfo{author}{Pedersen, N.~F.}
\newblock \bibinfo{title}{{Analysis of a Power Grid using a Kuramoto-like
  Model}}.
\newblock \emph{\bibinfo{journal}{{The European Physical Journal B}}}
  \textbf{\bibinfo{volume}{61}}, \bibinfo{pages}{485--491}
  (\bibinfo{year}{2008}).

\bibitem{Nishikawa2015}
\bibinfo{author}{Nishikawa, T.} \& \bibinfo{author}{Motter, A.~E.}
\newblock \bibinfo{title}{Comparative analysis of existing models for
  power-grid synchronization}.
\newblock \emph{\bibinfo{journal}{New Journal of Physics}}
  \textbf{\bibinfo{volume}{17}}, \bibinfo{pages}{015012}
  (\bibinfo{year}{2015}).

\bibitem{schmietendorf2017}
\bibinfo{author}{Schmietendorf, K.}, \bibinfo{author}{Peinke, J.} \&
  \bibinfo{author}{Kamps, O.}
\newblock \bibinfo{title}{The impact of turbulent renewable energy production
  on power grid stability and quality}.
\newblock \emph{\bibinfo{journal}{The European Physical Journal B}}
  \textbf{\bibinfo{volume}{90}}, \bibinfo{pages}{222} (\bibinfo{year}{2017}).

\bibitem{bottcher2019}
\bibinfo{author}{B{\"o}ttcher, P.~C.}, \bibinfo{author}{Otto, A.},
  \bibinfo{author}{Kettemann, S.} \& \bibinfo{author}{Agert, C.}
\newblock \bibinfo{title}{Time delay effects in the control of synchronous
  electricity grids}.
\newblock \emph{\bibinfo{journal}{arXiv preprint arXiv:1907.13370}}
  (\bibinfo{year}{2019}).

\bibitem{Haehne2018}
\bibinfo{author}{Haehne, H.}, \bibinfo{author}{Schottler, J.},
  \bibinfo{author}{Waechter, M.}, \bibinfo{author}{Peinke, J.} \&
  \bibinfo{author}{Kamps, O.}
\newblock \bibinfo{title}{The footprint of atmospheric turbulence in power grid
  frequency measurements}.
\newblock \emph{\bibinfo{journal}{Europhysics Letters}}
  \textbf{\bibinfo{volume}{121}}, \bibinfo{pages}{30001}
  (\bibinfo{year}{2018}).

\bibitem{Zhang2010}
\bibinfo{author}{Zhang, H.} \& \bibinfo{author}{Li, P.}
\newblock \bibinfo{title}{Probabilistic analysis for optimal power flow under
  uncertainty}.
\newblock \emph{\bibinfo{journal}{IET Generation, Transmission \&
  Distribution}} \textbf{\bibinfo{volume}{4}}, \bibinfo{pages}{553--561}
  (\bibinfo{year}{2010}).

\bibitem{Schaefer2017}
\bibinfo{author}{Sch{\"a}fer, B.} \emph{et~al.}
\newblock \bibinfo{title}{Escape routes, weak links, and desynchronization in
  fluctuation-driven networks}.
\newblock \emph{\bibinfo{journal}{Physical Review E}}
  \textbf{\bibinfo{volume}{95}}, \bibinfo{pages}{060203}
  (\bibinfo{year}{2017}).

\bibitem{Schaefer2017a}
\bibinfo{author}{Sch{\"a}fer, B.}, \bibinfo{author}{Beck, C.},
  \bibinfo{author}{Aihara, K.}, \bibinfo{author}{Witthaut, D.} \&
  \bibinfo{author}{Timme, M.}
\newblock \bibinfo{title}{Non-gaussian power grid frequency fluctuations
  characterized by l\'evy-stable laws and superstatistics}.
\newblock \emph{\bibinfo{journal}{Nature Energy}} \textbf{\bibinfo{volume}{3}}
  (\bibinfo{year}{2018}).

\bibitem{Weisbach2009}
\bibinfo{author}{Wei{\ss}bach, T.} \& \bibinfo{author}{Welfonder, E.}
\newblock \bibinfo{title}{High frequency deviations within the european power
  system: Origins and proposals for improvement}.
\newblock In \emph{\bibinfo{booktitle}{Power Systems Conference and Exposition,
  2009. PSCE'09. IEEE/PES}}, \bibinfo{pages}{1--6}
  (\bibinfo{organization}{IEEE}, \bibinfo{year}{2009}).

\bibitem{scoltock2015model}
\bibinfo{author}{Scoltock, J.}, \bibinfo{author}{Geyer, T.} \&
  \bibinfo{author}{Madawala, U.~K.}
\newblock \bibinfo{title}{Model predictive direct power control for
  grid-connected npc converters}.
\newblock \emph{\bibinfo{journal}{IEEE Transactions on industrial Electronics}}
  \textbf{\bibinfo{volume}{62}}, \bibinfo{pages}{5319--5328}
  (\bibinfo{year}{2015}).

\bibitem{dong2012frequency}
\bibinfo{author}{Dong, D.} \emph{et~al.}
\newblock \bibinfo{title}{Frequency behavior and its stability of
  grid-interface converter in distributed generation systems}.
\newblock In \emph{\bibinfo{booktitle}{2012 Twenty-Seventh Annual IEEE Applied
  Power Electronics Conference and Exposition (APEC)}},
  \bibinfo{pages}{1887--1893} (\bibinfo{organization}{IEEE},
  \bibinfo{year}{2012}).

\bibitem{tso2007predicting}
\bibinfo{author}{Tso, G.~K.} \& \bibinfo{author}{Yau, K.~K.}
\newblock \bibinfo{title}{Predicting electricity energy consumption: A
  comparison of regression analysis, decision tree and neural networks}.
\newblock \emph{\bibinfo{journal}{Energy}} \textbf{\bibinfo{volume}{32}},
  \bibinfo{pages}{1761--1768} (\bibinfo{year}{2007}).

\bibitem{sharma2011predicting}
\bibinfo{author}{Sharma, N.}, \bibinfo{author}{Sharma, P.},
  \bibinfo{author}{Irwin, D.} \& \bibinfo{author}{Shenoy, P.}
\newblock \bibinfo{title}{Predicting solar generation from weather forecasts
  using machine learning}.
\newblock In \emph{\bibinfo{booktitle}{2011 IEEE international conference on
  smart grid communications (SmartGridComm)}}, \bibinfo{pages}{528--533}
  (\bibinfo{organization}{IEEE}, \bibinfo{year}{2011}).

\bibitem{tchuisseu2017}
\bibinfo{author}{Tchuisseu, E.~T.}, \bibinfo{author}{Gomila, D.},
  \bibinfo{author}{Brunner, D.} \& \bibinfo{author}{Colet, P.}
\newblock \bibinfo{title}{Effects of dynamic-demand-control appliances on the
  power grid frequency}.
\newblock \emph{\bibinfo{journal}{Physical Review E}}
  \textbf{\bibinfo{volume}{96}}, \bibinfo{pages}{022302}
  (\bibinfo{year}{2017}).

\bibitem{Transnet}
\bibinfo{author}{{TransnetBW GmbH}}.
\newblock \bibinfo{title}{{Regelenergie Bedarf + Abruf}}
  (\bibinfo{year}{2019}).
\newblock
  \urlprefix\url{https://www.transnetbw.de/de/strommarkt/systemdienstleistungen/regelenergie-bedarf-und-abruf}.

\bibitem{Schafer2018b}
\bibinfo{author}{Sch\"afer, B.}, \bibinfo{author}{Timme, M.} \&
  \bibinfo{author}{Witthaut, D.}
\newblock \bibinfo{title}{Isolating the impact of trading on grid frequency
  fluctuations}.
\newblock In \emph{\bibinfo{booktitle}{2018 IEEE PES Innovative Smart Grid
  Technologies Conference Europe (ISGT-Europe)}}, \bibinfo{pages}{1--5}
  (\bibinfo{organization}{IEEE}, \bibinfo{year}{2018}).

\bibitem{Anvari2019}
\bibinfo{author}{Anvari, M.} \emph{et~al.}
\newblock \bibinfo{title}{Stochastic properties of the frequency dynamics in
  real and synthetic power grids}.
\newblock \emph{\bibinfo{journal}{arXiv preprint arXiv:1909.09110}}
  (\bibinfo{year}{2019}).

\bibitem{Machowski2011}
\bibinfo{author}{Machowski, J.}, \bibinfo{author}{Bialek, J.} \&
  \bibinfo{author}{Bumby, J.}
\newblock \emph{\bibinfo{title}{{Power System Dynamics: Stability and
  Control}}} (\bibinfo{publisher}{{John Wiley \& Sons,Chichester}},
  \bibinfo{year}{2011}).

\bibitem{oudalov2007optimizing}
\bibinfo{author}{Oudalov, A.}, \bibinfo{author}{Chartouni, D.} \&
  \bibinfo{author}{Ohler, C.}
\newblock \bibinfo{title}{Optimizing a battery energy storage system for
  primary frequency control}.
\newblock \emph{\bibinfo{journal}{IEEE Transactions on Power Systems}}
  \textbf{\bibinfo{volume}{22}}, \bibinfo{pages}{1259--1266}
  (\bibinfo{year}{2007}).

\bibitem{Wood2012}
\bibinfo{author}{Wood, A.~J.} \& \bibinfo{author}{Wollenberg, B.~F.}
\newblock \emph{\bibinfo{title}{Power generation, operation, and control}}
  (\bibinfo{publisher}{John Wiley \& Sons}, \bibinfo{year}{2012}).

\bibitem{Milano2018}
\bibinfo{author}{Milano, F.}, \bibinfo{author}{D{\"o}rfler, F.},
  \bibinfo{author}{Hug, G.}, \bibinfo{author}{Hill, D.~J.} \&
  \bibinfo{author}{Verbi{\v{c}}, G.}
\newblock \bibinfo{title}{Foundations and challenges of low-inertia systems}.
\newblock In \emph{\bibinfo{booktitle}{2018 Power Systems Computation
  Conference (PSCC)}}, \bibinfo{pages}{1--25} (\bibinfo{organization}{IEEE},
  \bibinfo{year}{2018}).

\bibitem{Beck2007}
\bibinfo{author}{Beck, H.-P.} \& \bibinfo{author}{Hesse, R.}
\newblock \bibinfo{title}{Virtual synchronous machine}.
\newblock In \emph{\bibinfo{booktitle}{Electrical Power Quality and
  Utilisation, 2007. EPQU 2007. 9th International Conference on}},
  \bibinfo{pages}{1--6} (\bibinfo{organization}{IEEE}, \bibinfo{year}{2007}).

\bibitem{Kundur2004}
\bibinfo{author}{Kundur, P.} \emph{et~al.}
\newblock \bibinfo{title}{Definition and classification of power system
  stability ieee/cigre joint task force on stability terms and definitions}.
\newblock \emph{\bibinfo{journal}{IEEE Transactions on Power Systems}}
  \textbf{\bibinfo{volume}{19}}, \bibinfo{pages}{1387--1401}
  (\bibinfo{year}{2004}).

\bibitem{Palensky2011}
\bibinfo{author}{Palensky, P.} \& \bibinfo{author}{Dietrich, D.}
\newblock \bibinfo{title}{{Demand Side Management: Demand Response, Intelligent
  Energy Systems, and Smart Loads}}.
\newblock \emph{\bibinfo{journal}{{IEEE Transactions on Industrial
  Informatics}}} \textbf{\bibinfo{volume}{7}}, \bibinfo{pages}{381--388}
  (\bibinfo{year}{2011}).

\bibitem{Kovacevic2013}
\bibinfo{author}{Kovacevic, R.~M.}, \bibinfo{author}{Pflug, G.~C.} \&
  \bibinfo{author}{Vespucci, M.~T.}
\newblock \emph{\bibinfo{title}{Handbook of risk management in energy
  production and trading}} (\bibinfo{publisher}{Springer},
  \bibinfo{year}{2013}).

\bibitem{NationalAcademiesofSciences2016}
\bibinfo{author}{{National Academies of Sciences Engineering and Medicine}}.
\newblock \emph{\bibinfo{title}{The Power of Change: Innovation for Development
  and Deployment of Increasingly Clean Electric Power Technologies}}
  (\bibinfo{publisher}{The National Academies Press, Washington, DC},
  \bibinfo{year}{2016}).

\bibitem{Weissbach2009}
\bibinfo{author}{Wei{\ss}bach, T.} \& \bibinfo{author}{Welfonder, E.}
\newblock \bibinfo{title}{High frequency deviations within the european power
  system--origins and proposals for improvement}.
\newblock \emph{\bibinfo{journal}{VGB powertech}}
  \textbf{\bibinfo{volume}{89}}, \bibinfo{pages}{26} (\bibinfo{year}{2009}).

\bibitem{Gardiner1985}
\bibinfo{author}{Gardiner, C.~W.}
\newblock \emph{\bibinfo{title}{Handbook of Stochastic Methods: for Physics,
  Chemistry and the Natural Sciences}} (\bibinfo{publisher}{Springer},
  \bibinfo{year}{1985}).

\bibitem{gonzalez2007forecasting}
\bibinfo{author}{Gonz{\'a}lez-Romera, E.},
  \bibinfo{author}{Jaramillo-Mor{\'a}n, M.~{\'A}.} \&
  \bibinfo{author}{Carmona-Fern{\'a}ndez, D.}
\newblock \bibinfo{title}{Forecasting of the electric energy demand trend and
  monthly fluctuation with neural networks}.
\newblock \emph{\bibinfo{journal}{Computers \& Industrial Engineering}}
  \textbf{\bibinfo{volume}{52}}, \bibinfo{pages}{336--343}
  (\bibinfo{year}{2007}).

\bibitem{Milan2013}
\bibinfo{author}{Milan, P.}, \bibinfo{author}{W{\"a}chter, M.} \&
  \bibinfo{author}{Peinke, J.}
\newblock \bibinfo{title}{{Turbulent Character of Wind Energy}}.
\newblock \emph{\bibinfo{journal}{{Physical Review Letters}}}
  \textbf{\bibinfo{volume}{110}}, \bibinfo{pages}{138701}
  (\bibinfo{year}{2013}).

\bibitem{Beck2003}
\bibinfo{author}{Beck, C.} \& \bibinfo{author}{Cohen, E. G.~D.}
\newblock \bibinfo{title}{Superstatistics}.
\newblock \emph{\bibinfo{journal}{Physica A}} \textbf{\bibinfo{volume}{322}},
  \bibinfo{pages}{267--275} (\bibinfo{year}{2003}).

\bibitem{Milan2014}
\bibinfo{author}{Milan, P.}, \bibinfo{author}{W{\"a}chter, M.} \&
  \bibinfo{author}{Peinke, J.}
\newblock \bibinfo{title}{Stochastic modeling and performance monitoring of
  wind farm power production}.
\newblock \emph{\bibinfo{journal}{Journal of Renewable and Sustainable Energy}}
  \textbf{\bibinfo{volume}{6}}, \bibinfo{pages}{033119} (\bibinfo{year}{2014}).

\bibitem{Beck2005}
\bibinfo{author}{Beck, C.}, \bibinfo{author}{Cohen, E. G.~D.} \&
  \bibinfo{author}{Swinney, H.~L.}
\newblock \bibinfo{title}{From time series to superstatistics}.
\newblock \emph{\bibinfo{journal}{Physical Review E}}
  \textbf{\bibinfo{volume}{72}}, \bibinfo{pages}{056133}
  (\bibinfo{year}{2005}).

\bibitem{Ulbig2014}
\bibinfo{author}{Ulbig, A.}, \bibinfo{author}{Borsche, T.~S.} \&
  \bibinfo{author}{Andersson, G.}
\newblock \bibinfo{title}{Impact of low rotational inertia on power system
  stability and operation}.
\newblock \emph{\bibinfo{journal}{IFAC Proceedings Volumes}}
  \textbf{\bibinfo{volume}{47}}, \bibinfo{pages}{7290--7297}
  (\bibinfo{year}{2014}).

\bibitem{Weitenberg2017}
\bibinfo{author}{Weitenberg, E.} \emph{et~al.}
\newblock \bibinfo{title}{Robust decentralized secondary frequency control in
  power systems: Merits and trade-offs}.
\newblock \emph{\bibinfo{journal}{arXiv preprint arXiv:1711.07332}}
  (\bibinfo{year}{2017}).

\bibitem{tchuisseu2018curing}
\bibinfo{author}{Tchuisseu, E. B.~T.} \emph{et~al.}
\newblock \bibinfo{title}{Curing braess’ paradox by secondary control in
  power grids}.
\newblock \emph{\bibinfo{journal}{New Journal of Physics}}
  \textbf{\bibinfo{volume}{20}}, \bibinfo{pages}{083005}
  (\bibinfo{year}{2018}).

\bibitem{wang2016control}
\bibinfo{author}{Wang, C.}, \bibinfo{author}{Grebogi, C.} \&
  \bibinfo{author}{Baptista, M.~S.}
\newblock \bibinfo{title}{Control and prediction for blackouts caused by
  frequency collapse in smart grids}.
\newblock \emph{\bibinfo{journal}{Chaos: An Interdisciplinary Journal of
  Nonlinear Science}} \textbf{\bibinfo{volume}{26}}, \bibinfo{pages}{093119}
  (\bibinfo{year}{2016}).

\bibitem{Simpson-Porco2012}
\bibinfo{author}{Simpson-Porco, J.~W.}, \bibinfo{author}{D{\"o}rfler, F.} \&
  \bibinfo{author}{Bullo, F.}
\newblock \bibinfo{title}{Droop-controlled inverters are kuramoto oscillators}.
\newblock \emph{\bibinfo{journal}{IFAC Proceedings Volumes}}
  \textbf{\bibinfo{volume}{45}}, \bibinfo{pages}{264--269}
  (\bibinfo{year}{2012}).

\bibitem{hammid2016load}
\bibinfo{author}{Hammid, A.~T.}, \bibinfo{author}{Hojabri, M.},
  \bibinfo{author}{Sulaiman, M.~H.}, \bibinfo{author}{Abdalla, A.~N.} \&
  \bibinfo{author}{Kadhim, A.~A.}
\newblock \bibinfo{title}{Load frequency control for hydropower plants using
  pid controller}.
\newblock \emph{\bibinfo{journal}{Journal of Telecommunication, Electronic and
  Computer Engineering (JTEC)}} \textbf{\bibinfo{volume}{8}},
  \bibinfo{pages}{47--51} (\bibinfo{year}{2016}).

\bibitem{Dongmo2017}
\bibinfo{author}{Dongmo, E.~D.}, \bibinfo{author}{Colet, P.} \&
  \bibinfo{author}{Woafo, P.}
\newblock \bibinfo{title}{Power grid enhanced resilience using proportional and
  derivative control with delayed feedback}.
\newblock \emph{\bibinfo{journal}{The European Physical Journal B}}
  \textbf{\bibinfo{volume}{90}}, \bibinfo{pages}{6} (\bibinfo{year}{2017}).

\bibitem{martyr2019benchmarking}
\bibinfo{author}{Martyr, R.}, \bibinfo{author}{Sch{\"a}fer, B.},
  \bibinfo{author}{Beck, C.} \& \bibinfo{author}{Latora, V.}
\newblock \bibinfo{title}{Benchmarking the performance of controllers for power
  grid transient stability}.
\newblock \emph{\bibinfo{journal}{Sustainable Energy, Grids and Networks}}
  \textbf{\bibinfo{volume}{18}}, \bibinfo{pages}{100215}
  (\bibinfo{year}{2019}).

\bibitem{ENTSOE_Demand}
\bibinfo{author}{{ENTSO-E}}.
\newblock \bibinfo{title}{{ Generation Forecast - Day ahead}}.
\newblock \bibinfo{howpublished}{{ENTSO-E}
  \url{https://transparency.entsoe.eu/generation/r2/dayAheadAggregatedGeneration/show}}
  (\bibinfo{year}{2019}).

\bibitem{Risken1984a}
\bibinfo{author}{Risken, H.}
\newblock \emph{\bibinfo{title}{The Fokker-Planck Equation}}
  (\bibinfo{publisher}{Springer, Berlin}, \bibinfo{year}{1984}).

\bibitem{friedrich2011}
\bibinfo{author}{Friedrich, R.}, \bibinfo{author}{Peinke, J.},
  \bibinfo{author}{Sahimi, M.} \& \bibinfo{author}{Tabar, M. R.~R.}
\newblock \bibinfo{title}{Approaching complexity by stochastic methods: From
  biological systems to turbulence}.
\newblock \emph{\bibinfo{journal}{Physics Reports}}
  \textbf{\bibinfo{volume}{506}}, \bibinfo{pages}{87--162}
  (\bibinfo{year}{2011}).

\bibitem{NationalGrid2019}
\bibinfo{author}{{National Grid ESO}}.
\newblock \bibinfo{title}{Historic frequency data} (\bibinfo{year}{2019}).
\newblock
  \urlprefix\url{https://www.nationalgrideso.com/balancing-services/frequency-response-services/historic-frequency-data}.

\bibitem{Gorjao2019b}
\bibinfo{author}{Gorj\~ao, L.~R.} \& \bibinfo{author}{Meirinhos, F.}
\newblock \bibinfo{title}{{P}ython {KM}: {Kramers--Moyal} coefficients for
  stochastic processes}.
\newblock \emph{\bibinfo{journal}{Journal of Open Source Software, under
  review}}  (\bibinfo{year}{2019}).

\bibitem{Rinn16}
\bibinfo{author}{Rinn, P.}, \bibinfo{author}{Lind, P.~G.},
  \bibinfo{author}{Wächter, M.} \& \bibinfo{author}{Peinke, J.}
\newblock \bibinfo{title}{The {L}angevin approach: An {R} package for modeling
  markov processes}.
\newblock \emph{\bibinfo{journal}{Journal of Open Research Software}}
  \textbf{\bibinfo{volume}{4}}, \bibinfo{pages}{e34} (\bibinfo{year}{2016}).

\end{thebibliography}

\newpage ~ \newpage

\onecolumngrid
\section{Supplemental Material}
\subsection{Parameter extraction guidelines}\label{Sec:guidelines}
Following the mathematical foundations presented in the main text,%previous section, 
we present hands-on instructions on how to extract the parameters for the example of a month-long recording of the power-grid frequency in Germany for the month of January 2019.
As we focus mainly on specific characteristics of the power dynamics, we calculate, strictly from the data, the noise amplitude $\epsilon$, the power mismatch at the hourly stamp $\Delta P$, the primary and secondary control amplitudes $c_1$ and  $c_2$.
The procedure follows in a simple manner:
\begin{itemize}
    \item Noise amplitude $\epsilon$: Utilise the second Kramers--Moyal coefficient, i.e., the diffusion, to extract the noise strength $\epsilon$ from the timeseries of the data.
    Use relation \eqref{eq:Diffusion_Coefficient} to obtain the value, by taking either the value of the diffusion at $f=50$ Hz or averaging in windows around $f=50$ Hz.
    \item Power mismatch $\Delta P$ (for the hourly jumps): Take the first $10$ seconds of data just after the hour, e.g. from 12:00:00 to 12:00:10. Calculate the slope of the frequency increase or decrease in this window with a linear fit.
    Given that the process displays jumps up and down, i.e., excess and lack of power supply, take the absolute value to obtain the general power mismatch $\Delta P$.
    Average to obtain the average effect.
    \item Primary control $c_1$: This is a two-step process: Perform a Gaussian kernel de-trending of the data, with a $60$-seconds window, to remove the effects of the market and dispatch, so to capture the system's stochastic nature.
    The choice of a $60$-second window ensures one removes only the deterministic characteristics of the frequency trajectory: a smaller window will mimic the noise, a larger window will reflect the overall mean of $50$~{Hz} ($60$~{Hz}) of the process.
    Utilise now the first Kramers--Moyal coefficient, i.e., the drift term, to obtain a negatively tilted line: linearly fit the line around $f=50$~{Hz} (or $60$~{Hz}) and extract the slope, which is the drift coefficient of the governing Ornstein--Uhlenbeck process.
    The slope is the negative primary control $-c_1$.
    \item Secondary control $c_2$: This is the last parameter to calculate, and it depends on the primary control $c_1$. Take $900$ seconds windows at every hourly jump, similarly to the above calculations for the power mismatch $\Delta P$.
    Fit \eqref{eq:approx_of_7} to the data snippets (or \eqref{eq:OmegaSolution_afterJump}, although strictly mathematically correct, it is harder to fit).
    Obtain the exponential decay made explicit in \eqref{eq:exp-decay-omega}, i.e., the last term of \eqref{eq:approx_of_7}.
    Input the previously obtained value for the primary control $c_1$ (step above) to determine teh secondary control $c_2$.
\end{itemize}

Having concluded these four steps, we possess all the necessary variables to numerically integrate a synthetic version of the evaluated power-grid frequency.

The simplest and most straightforward method is to implement an Euler--Mayurama integration scheme. This is a scheme identical to a regular Euler integration scheme, incorporating a noise function $\xi$.
This is done by generating a set of normally distributed values with mean $\mu=0$ and variance $\sigma=\sqrt{\tau}$, with $\tau$ the employed time-step of integration.
Stochastic integration requires small time-steps, thus we suggest using at least $0.01$ seconds, or better even $0.001$ seconds.
From this store only the $1$ second recording to accurately compare with available real power-grid data  (if your temporal resolution is different, match it).
Other more integrators, such as Runge-Kutta integrators for stochastic equations, can be used to ensure higher precision of the numerical results.

To extract the Kramers--Moyal coefficients there are open source Python (`Python KM') or R (`Langevin') packages, see \cite{Gorjao2019b} and \cite{Rinn16}, respectively.

\section{Pseudo-code}

Pseudo-code for extracting the parameters from data, based on the methodology implemented for the Central European power grid.
As Supplemental Material, a minimal \texttt{python} code is attached.
This was the code used for obtaining the parameters from the data.

In the following we compartmentalise the code in four sections, each corresponding to the parameter recovery of each of the four parameters under analysis: Noise $\epsilon$, primary control $c_1$, secondary control $c_2$, and dispatch $\Delta P$

For all cases below, the first step is naturally to import the data

\textbf{Import data}
\begin{itemize}
	\item Load \texttt{data}
	\begin{itemize}
		\item[IF] \texttt{data} is recorded at 50 hz: \texttt{data}~$=$~\texttt{data}~$-~50$
	\end{itemize}
\end{itemize}~
\\[2em]~
Retrieving the Noise $\epsilon$
\begin{itemize}
	\item Load module \texttt{km} to obtain Kramers--Moyal coefficients
	\item \texttt{diffusion}, \texttt{space}~$=$~\texttt{km}(\texttt{data}, coefficient $= 2$)
	\item find $f\!=\!0$ in \texttt{space}
	\item $\epsilon = \sqrt{\text{\texttt{diffusion}(\texttt{space}~$=0$)} \times 2}$
\end{itemize}~
\\[2em]~
\textbf{Retrieving the primary control $c_1$}
\begin{itemize}
	\item Load module \texttt{km} to obtain Kramers--Moyal coefficients
	\item Load module \texttt{filter} to obtain the Gaussian kernel filtering
	\item \texttt{data\_filtered}~$=$~\texttt{data}~$-$~ \texttt{filter}(\texttt{data})
	\item \texttt{drift}, \texttt{space}~$=$~\texttt{km}(\texttt{data\_filtered}, coefficient $= 1$)
	\item find $f\!=\!0$ in \texttt{space}
	\item fit line to \texttt{drift} around \texttt{space}~$=0$
	\item $c_1 = - \text{slope of fit}$ 
\end{itemize}~
\\[2em]~

\textbf{Retrieving the dispatch $\Delta P$}
\begin{itemize}
	\item[FOR] every hour:
	\begin{itemize}
		\item fit line to \texttt{data}[first 10 secs]
		\item save slope to \texttt{record}
	\end{itemize}
	\item Take absolute  of \texttt{record}
	\item $\Delta P = $ mean(abs(\texttt{record}))
\end{itemize}~
\\[2em]~
\textbf{Retrieving the secondary control $c_2$}
\begin{itemize}
	\item[FOR] every hour
	\begin{itemize}
		\item fit curve of \eqref{eq:approx_of_7}
		to \texttt{data}[900 seconds]
		\item save exp. decay to \texttt{record}
	\end{itemize}
	\item $c_2 = $ mean(\texttt{record})$\times c_1$ 
\end{itemize}
It is advisable to discard the statistical outliers, since fitting an exponential decay to the frequency data is especially unreliable if the dispatch difference is very small for that period.

\section{Python minimal-working code}
\begin{lstlisting}[language=Python, caption=Load libraries and data]
# Set of required python libraries
import numpy as np
from scipy.optimize import curve_fit

# Library for the gaussian kernel filter
from scipy.ndimage.filters import gaussian_filter1d

# Library for calculating Kramers--Moyal coefficients
from kramersmoyal import km


# Preliminaries
# Allocate the power-grid frequency data to a numpy array. Make sure the first
# entry corresponds to the zero second of an hour period, e.g. data[0] is the
# start of the data at some HH:00:00

data = np.readtxt('location/of/data.txt')

# if the data is recorded at a reference (e.g. 50 Hz), remove the reference
data = data - 50.0
\end{lstlisting}

\begin{lstlisting}[language=Python, caption=Noise $\epsilon$]
# Noise epsilon
# In order to calculate the noise epsilon you need to extract the diffusion term
# of the stochastic processes. Employ the km function from the kramersmoyal
# library

# Retrieve the diffusion coefficient
diffusion, space = km(data, powers = [0, 2], bins = np.array([6000]), bw = 0.05)

# find the zero frequency
zero_frequency = np.argmin(space[0]**2)

# evaluate the diffusion at that point and extract epsilon
epsilon = np.sqrt(diffusion[1,zero_frequency]*2)
\end{lstlisting}

\begin{lstlisting}[language=Python, caption=Primary control $c_1$]
# Primary control c_1
# To calculate the primary control c_1 we need to emply a two step process.
# First remove the general trend by a gaussian kernel filtering, then employ
# again the km function from the kramersmoyal librayr to obtain the drift term

data_filter = gaussian_filter1d(data, sigma = 60)

# Obtain the drift coefficient
drift, space = km(data-data_filter, powers = [0, 1], bins = np.array([6000]), bw = 0.01)

# find the zero frequency
mid_point = np.argmin(space[0]**2)

# Calculate the slope of the drift term, which gives the primary control c_1.
# The fiting is to a line of intercept a and slope b. T
c_1 = curve_fit(lambda t,a,b: a - b*t,  space[0][mid_point - 500:mid_point + 500],
        drift[1,mid_point - 500:mid_point + 500] ,  p0=(0.0002, 0.005),
        maxfev=10000
        )[0][1]
\end{lstlisting}

\begin{lstlisting}[language=Python, caption=$\Delta P$]
# Delta P / RoCoF
# To calculate the dispatch Delta P evaluate the process at every hourly jump.
# If there is a different dispatch seems, change the evaluation to that period.
# In principle this can be calculated for any interval of power dispatch, but
# to ensure a good fit, bigger jumps = bigger dispatch = better fit

# Define window of jumps. In this case, evaluate the Delta P every hour
window = 3600   # 3600 seconds = 1 hour

# Set the total length to evaluate
data_range = data.size // window

# Initialise an array to record the Delta P
Delta_P_slopes = np.zeros(data_range)

# The jumps are to be evaluate at t=0, but since we have noise data, we fit the
# first 10 seconds to calculate the slope
for j in range(data_range):
    Delta_P_slopes[j] = curve_fit(lambda t,a,b: a + b*t,  np.linspace(0,9,10),
                        data[3600*(j):3600*(j)+10],  p0=(0.0, 0.0),
                        maxfev=10000
                        )[0][1]
# This results is an array with positive and negative slopes, since some
# frequency changes are positive (excess energy), some are negative. Find the
# absolute value for them and take the average as the reference Delta P.

# This is the mean Delta P
Delta_P = np.mean(np.abs(Delta_P_slopes))
\end{lstlisting}

\begin{lstlisting}[language=Python, caption=Secondary control $c_2$]
# Secondary control c_2
# To calculate the secondary control c_2 we will need, just as above, snippets
# of the hourly jumps and the subsequent decay of the frequency back to the
# nominal values. Due to the complicated frequency behaviour, we will fit an
# entire curve to the 900 seconds but we shall only extract the decay rate

# Define window of jumps. In this case, evaluate the secondary control c_2 every
# hour
window = 3600   # 3600 seconds = 1 hour

# Set the total length to evaluate
data_range = data.size // window

# Initialise an array to record the Delta P
c_2_decays = np.zeros(data_range)

# Since we have up and down jumps, we have to separate the trajectories that
# move up and those that move down, but we still calculate the same decay
# behaviour of both
for j in range(data_range):
    # if the frequency trajectory moves positively
    if np.sum((np.diff(data[3600*(j):3600*(j)+10]))) > 0:
        c_2_decays[j] = curve_fit(lambda t,a,b,c:
            a*np.exp(-b*t)*(1-np.exp(-c*t+2*b*t)),
            np.linspace(0,899,900), data[3600*(j):3600*(j)+900],
            p0=(0.08, .0045, 0.035), maxfev=10000
            )[0][1]
    else:
        c_2_decays[j] = curve_fit(lambda t,a,b,c:
            -a*np.exp(-b*t)*(1-np.exp(-c*t+2*b*t)),
            np.linspace(0,899,900), data[3600*(j):3600*(j)+900],
            p0=(0.08, .0045, 0.035), maxfev=10000
            )[0][1]

# We have thus stored the decay rate b of every power mismatch in the system.
# Due to statistical outliers, discard 20% of the data

# Sort the array and discard 20% of the largest values
temp_c_2_decays = c_2_decays[np.argsort(c_2_decays)][:-c_2_decays.size//5]

# Recall here that to calculate c_2 you need to know c_1
c_2 = np.mean(temp_c_2_decays) * c_1
\end{lstlisting}

\begin{lstlisting}[language=Python, caption=Print results]

# Print out results:
print(r' epsilon  |    c_1    |    c_2    |  Delta P   ')
print(r'----------|-----------|-----------|------------')
print(r'{0:.5f}'.format(epsilon,1) + '  |  '
    +  r'{0:.5f}'.format(c_1,1) + '  |  '
    +  r'{0:.5f}'.format(c_2,1) + '  |  '
    +  r'{0:.5f}'.format(Delta_P,1)
     )
\end{lstlisting}
\end{document}